\documentstyle[12pt]{article}

\title{First Principles Investigation of Ferromagnetism and
Ferroelectricity in Bismuth Manganite}
\author{Nicola A. Hill\\
Materials Department, University of California\\
Santa Barbara, CA 93106-5050\\
and\\
 Karin M. Rabe\\
Department of Applied Physics,Yale University\\P.O. Box 208284,
New Haven, CT 06520-8284}

\date{\today}

\begin{document}

\input psfig
\maketitle

\begin{abstract}

We present results of local spin density approximation (LSDA)
pseudopotential calculations for the perovskite structure oxide,
bismuth manganite (BiMnO$_3$). The origin of the differences between 
bismuth manganite and other perovskite manganites is determined by 
first calculating total energies and band structures of the high 
symmetry cubic phase, then 
sequentially lowering the magnetic and structural symmetry. 
Our results indicate that covalent bonding between bismuth
cations and oxygen anions stabilizes different magnetic and 
structural phases compared with the rare earth manganites.
This is consistent with recent experimental results showing
enhancement of charge ordering in doped bismuth manganite.

\end{abstract}

\section{Introduction}

\begin{figure}
\centerline{\psfig{figure=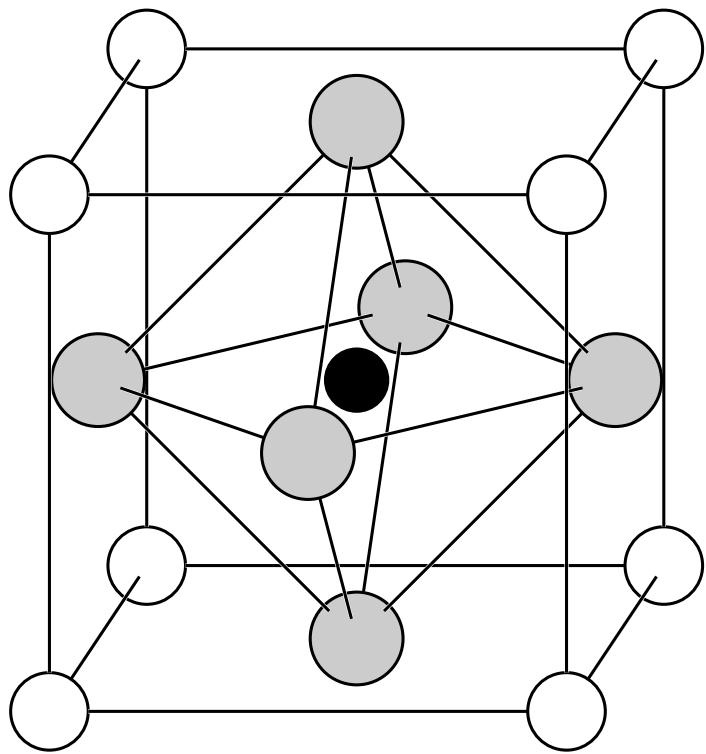,height=5in,width=5in}}
\caption{The perovskite structure. The small B cation (in black) is at
the center of an octahedron of oxygen anions (in gray). The large A cations 
(white) occupy the unit cell corners.}
\label{perovskite}
\end{figure}

Perovskite structure oxides exhibit a wide range of low temperature 
structural distortions associated with lattice instabilities of the 
prototype cubic structure (Figure~\ref{perovskite}), including
ferroelectric (e.g. PbTiO$_3$ and BaTiO$_3$), antiferroelectric (e.g.
PbZrO$_3$) and antiferrodistortive (e.g. SrTiO$_3$). Both first- and
second-order transitions are observed, with a full spectrum of
transition behavior ranging from displacive to order-disorder.
The strong coupling between the electric polarization and the 
structural distortions has led to the widespread use of perovskites 
such as lead
zirconium titanate (PZT) in piezoelectric transducer and actuator 
applications\cite{Cross}.

In {\it magnetic} perovskites, different microscopic orientations of the
spin-polarized ions give rise to different macroscopic magnetic symmetries.
This results in
rich phase diagrams, in which both the magnetic order and the structure
depend
on the temperature, pressure  and chemical composition.
A large amount of recent research activity \cite{Atlanta} has been
focused on the rare earth perovskite-structure manganites, following 
the observation of
colossal magnetoresistance (CMR) in Ca-doped LaMnO$_3$\cite{Jin}.
During these recent studies,
many rare earth perovskite manganites have been found to show strong
coupling between their magnetic  and structural, or magnetic and
electronic order parameters. For example,
a magnetically induced structural phase transition has been observed in
La$_{0.83}$Sr$_{0.17}$MnO$_3$\cite{Asamitsu} indicating strong coupling
between the local magnetic spin moments and the lattice structure.
In Nd$_{0.5}$Sr$_{0.5}$MnO$_3$, strong coupling
between the magnetic spin moments and the electronic charge
carriers was demonstrated when an electronic metal-insulator transition
was induced by an external magnetic field\cite{Kuwahara}.
The large change in conductivity with applied magnetic field,
which gives these materials potential technological importance as
the read element in magnetic recording heads\cite{Brug},
is believed to originate from a similar type of phase 
transition\cite{Kuwahara2}.

Materials which have  strong coupling between {\it all three} of the 
magnetic, electric and structural order parameters, resulting in 
simultaneous ferromagnetism, ferroelectricity and ferroelasticity,
are known as multiferroics. Most multiferroic
materials are complex structures with many atoms per formula
unit, and more than one formula unit per unit cell. The large
number of inter-ionic interactions has prevented isolation of
the essential factors causing multiferroicity. As a result,
the origin of multiferroism and the nature of the coupling
between the magnetic, electric polarization, and structural order 
parameters are not well understood. 
In spite of the lack of fundamental understanding, a number of 
device applications have been suggested for multiferroic
materials including multiple state memory elements, electric
field controlled ferromagnetic resonance devices, and variable
transducers with either magnetically modulated piezoelectricity
or electrically modulated piezomagnetism\cite{Wood_Austin}.
Also, the ability to couple with {\it either} the magnetic {\it or} 
the electric polarization offers an extra degree of freedom in the 
design of conventional actuators, transducers and storage devices.

Bismuth manganite can be regarded as the ``hydrogen atom'' of
multiferroics. Although information
about BiMnO$_3$ is sparse\cite{BiMnO3}, indications are that it
is simultaneously ferromagnetic and ferroelectric at low temperatures,
and, because it has a simple structure, it is amenable
to detailed study using first-principles techniques.
Investigation of BiMnO$_3$ permits the study of the
essentials of multiferroism without the problems
associated with simulating compounds containing many
atoms per formula unit.                                 
In addition, BiMnO$_3$ is strikingly different,
both magnetically and structurally, from the rare earth perovskite manganites.
BiMnO$_3$ is ferromagnetic with a triclinic structural distortion in
its ground state, whereas the rare earth manganites are antiferromagnetic
and either orthorhombic or hexagonal. The differences are unexpected
because bismuth and the rare earths all form trivalent cations
with similar ionic radii.
The anomalous behavior of BiMnO$_3$ compared with 
conventional perovskite manganites could give rise to unusual and 
useful transport properties.

The goal of the study described in this paper is to determine the origin of
the differences between BiMnO$_3$ and the other perovskite manganites,
both to understand the fundamental physics, and to assist in
the optimization of perovskite manganite materials for novel
device applications. To achieve this goal we evaluate electronic
and magnetic properties of perovskite manganites using a plane
wave pseudopotential (PWPP) implementation of density functional theory
(DFT) within the local spin density approximation (LSDA).
Our choice of theoretical approach is influenced by
successful first-principles studies of {\it non-magnetic} perovskite 
ferroelectrics\cite{Waghmare_Rabe,Cohen_Krakauer}. The origin of 
ferroelectricity in the prototypical ferroelectric perovskites
PbTiO$_3$ and BaTiO$_3$, and the reasons for the differences
between them, were deduced using first principles density
functional theory (DFT) calculations\cite{Cohen,Cohen_Krakauer}.
It was shown in both cases that Ti $3d$ - O $2p$ hybridization
is essential for ferroelectricity. In BaTiO$_3$, the
Ba$^{2+}$ ions are almost entirely ionically bonded, resulting
in a rhombohedral structure. In PbTiO$_3$, however, the Pb $6s$
electrons are partially covalently bonded with the oxygen $2p$
electrons, favoring a tetragonal structure.

Throughout this work, we use LaMnO$_3$ as our example of a typical rare 
earth manganite, and compare our calculated results for BiMnO$_3$ with 
those for LaMnO$_3$.  Since
Bi is adjacent to Pb, near the right hand side of the periodic 
table, and La is adjacent to Ba on the left hand side, we anticipate
by analogy with the earlier PbTiO$_3$-BaTiO$_3$ study, 
that the high lying occupied Bi $6s$ electrons will contribute to the 
structural and magnetic differences between BiMnO$_3$ and LaMnO$_3$.
In addition we will find significant hybridization between the Bi $6p$
and O $2p$ electrons. 

The remainder of this paper is organized as follows:
In Section ~\ref{theorysection} we describe the plane wave pseudopotential 
LSDA implementation of density functional theory
which we use for this study. Our PWPP implementation is
shown to give results which are in excellent agreement with
previously published all electron calculations for magnetic
perovskite oxides. In Section~\ref{Results} our results for
BiMnO$_3$ and LaMnO$_3$ are presented. Our results are
summarized in Section~\ref{Summary}.

\section{Computational Technique}
\label{theorysection}

The calculations described in this work were performed using
a plane wave pseudopotential implementation\cite{Yin} of density functional 
theory\cite{HKKS} within the local spin density approximation. 
Plane wave basis sets offer many advantages in total
energy calculations for solids, including completeness,
an unbiased representation, and arbitrarily good
convergence accuracy. They also allow for straightforward
mathematical formulation and implementation, which is
essential in the calculation of Hellmann-Feynman forces\cite{Bendt}
and in density functional perturbation theory calculations\cite{GonzePT}.

However plane wave basis sets necessitate the use of
pseudopotentials to model the electron-ion interaction,
in order to avoid rapid oscillations of the valence
wavefunctions in the region around the ion cores. The
accuracy and efficiency of {\it ab initio} pseudopotential
calculations (compared with all electron calculations) is now
well established for non-magnetic systems\cite{MLCohen}, but
application 
of the method to spin-polarized magnetic systems is still in its
infancy. Traditionally, magnetic materials containing transition
metals have been studied using all electron methods with mixed 
basis sets, such as the
linear augmented plane wave (LAPW)\cite{Singh}, linear muffin tin 
orbital\cite{LMTO}
or Korringa-Kohn-Rostoker\cite{KKR} approaches. 

Resistance to the
use of the PWPP method to study these systems has been based on two
perceived difficulties: 

1) First, the magnetic properties of the transition metals and
their compounds originate from tightly bound $d$-electrons, which 
might be expected to require a prohibitively large number of plane waves to 
expand their pseudopotentials. This problem can be overcome by
using the optimized pseudopotentials developed by Rappe et al.\cite{Rappe}. 
Optimized pseudopotentials minimize the kinetic energy
in the high Fourier components of the pseudo wavefunction,
leading to a corresponding reduction in the contribution of 
high Fourier components in the solid. The use of optimized
pseudopotentials reduces the energy cut-off of the plane wave
expansion in transition metal solids to around 60 Ry or less. 

2) Second, implicit within the pseudopotential approximation is
the assumption that the exchange-correlation potential, $V_{xc}$, is 
separable into a valence part and a core part:
\begin{equation}
V_{xc}(\rho,\sigma) = 
\left[ V_{xc}(\rho,\sigma) - V_{xc}(\rho^v,\sigma^v) \right ]
+ V_{xc}(\rho^v,\sigma^v) .
\label{Vxc}
\end{equation}
Here $\rho$ is the electron density, $\sigma$ is the spin polarization,
and the superscript $v$ refers to the contribution from the valence
electrons. If the core electrons are spatially separated from the
valence electrons then this separation is rigorously correct.
However in transition metals, the $d$-electrons have large wavefunction 
amplitudes near the core region. Therefore, because $V_{xc}$ is a non-linear
function of both charge and spin polarization, use of Eqn.~\ref{Vxc} is
inaccurate.  Early pseudopotential calculations for magnetic materials
attempted to circumvent this problem by using a small core 
radius \cite{Greenside_Schluter,Zhu}. This resulted in a 
pseudo-wavefunction which was similar to the all electron wavefunction and 
therefore could not be expanded efficiently using plane waves.
A better solution is offered by the partial 
non-linear core correction scheme of Louie et al.\cite{NLCCs}, in which the
core charge density is explicitly retained in both the construction of 
the ionic pseudopotential, and in the DFT calculation.

The PWPP method with optimized pseudopotentials and partial 
non-linear core corrections was first tested by Sasaki et al. 
for ferromagnetic nickel and iron\cite{Sasaki}. The structural 
and magnetic properties calculated using the PWPP method were 
found to be in excellent agreement with results from all electron 
calculations.  A similar study using different soft pseudopotentials 
obtained comparable results for the ground-state properties of 
nickel\cite{Cho_Kang}. More recently, Lewis and coworkers presented 
the first PWPP calculation for a magnetic {\it compound} 
(CrO$_2$)\cite{Lewis}, and showed that the results agreed well with 
those of all electron LSDA calculations. Before presenting our results 
for BiMnO$_3$ in Section~\ref{Results}, we will compare the results of 
our PWPP calculations for a magnetic {\it perovskite} with published 
all-electron calculations\cite{Pickett_Singh} and show that similarly 
good agreement is obtained.

\subsection{Pseudopotential Construction}

\begin{figure}
\centerline{\psfig{figure=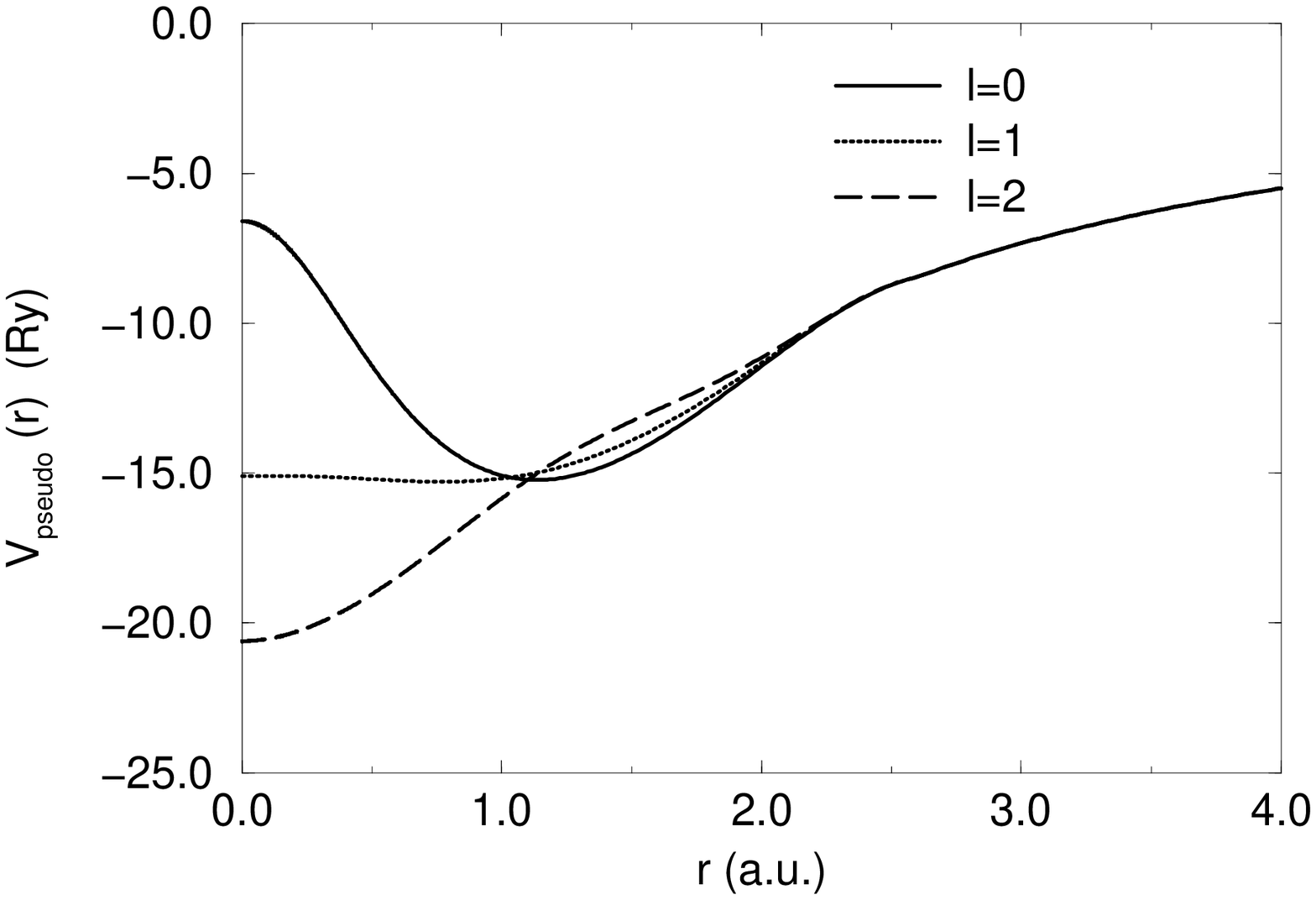,height=5.5in,width=6.0in}}
\caption{Optimized ionic pseudopotentials for La. The core radius is 2.6 a.u.
for all angular momenta. Four basis functions were included in the optimization,
which used a cutoff wavevector of 7.5 a.u.}
\label{La_PP}
\end{figure}

\begin{figure}
\centerline{\psfig{figure=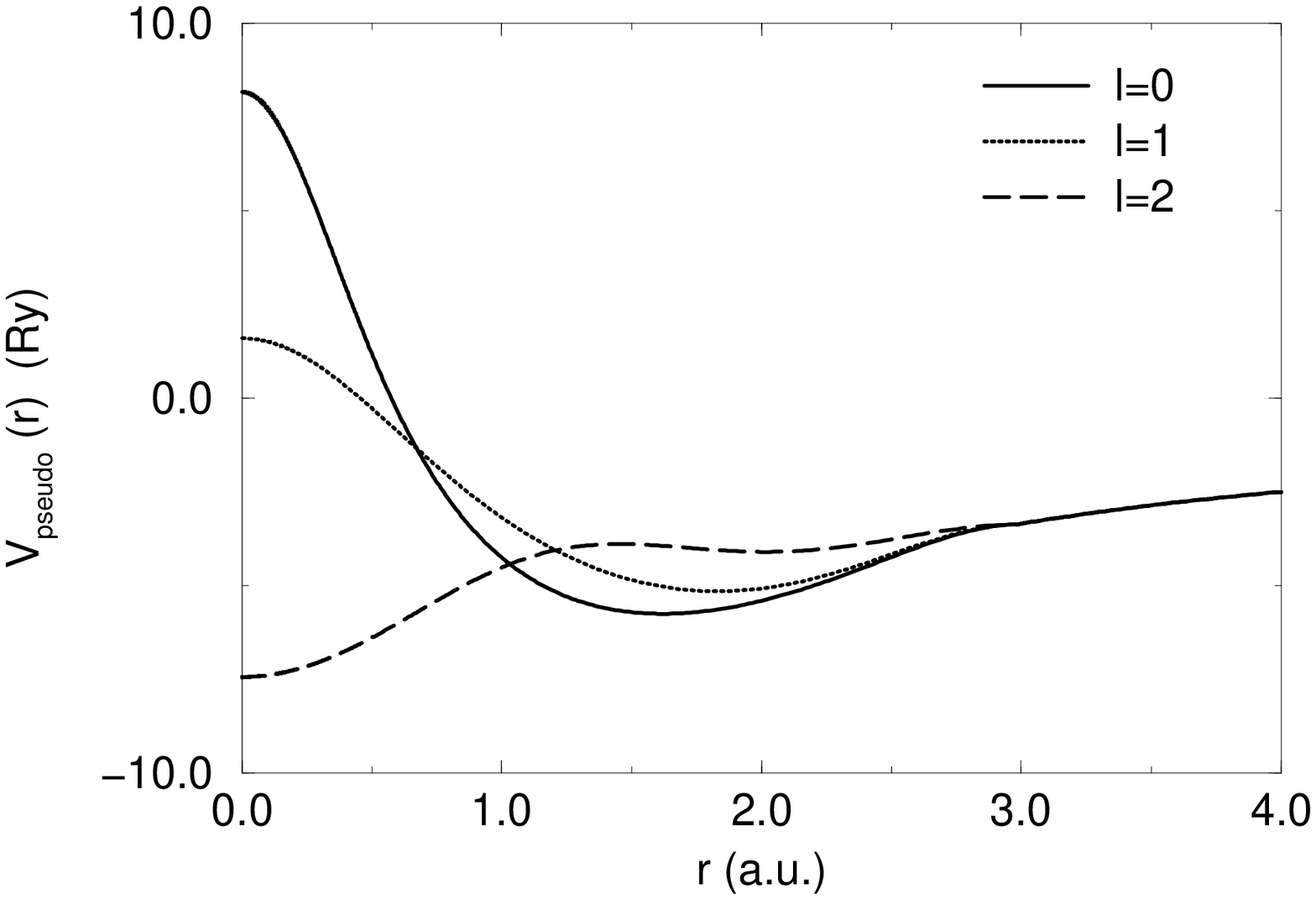,height=5.5in,width=6.0in}}
\caption{Optimized ionic pseudopotentials for Bi. The core radius is 3.0 a.u.
for all angular momenta. Four basis functions were included in the optimization,
which used a cutoff wavevector of 7.5 a.u.}
\label{Bi_PP}
\end{figure}

For La and Bi, we constructed optimized scalar-relativistic 
pseudopotentials, using (for La) a $5s^2 5p^6 5d^1$ reference 
configuration with core radius ($r_c$) $= 2.6$ a.u., and (for Bi)
a $6s^{0.5} 6p^{1.2} 6d^{0.3}$ reference configuration with $r_c = 3.0$ 
a.u. In the La reference configuration the 4$f$ orbitals were unoccupied,
and for Bi the filled 4$f$ shell was placed in the core. The pseudopotentials 
were optimized using 4 basis functions with a cutoff wavevector, $q_c$, 
of $7.5$ a.u. $q_c$ determines the convergence of the kinetic energy 
with respect to the plane wave cutoff energy in reciprocal space 
calculations.  The resulting optimized pseudopotentials are shown in
Figures~\ref{La_PP} and ~\ref{Bi_PP}.  The pseudopotentials were
tested for transferability by comparing with all electron calculations
for a range of typical atomic and ionic configurations. The 
pseudo-eigenvalues and total energies are equivalent to the all electron
values to within a few meVs.

\begin{figure}
\centerline{\psfig{figure=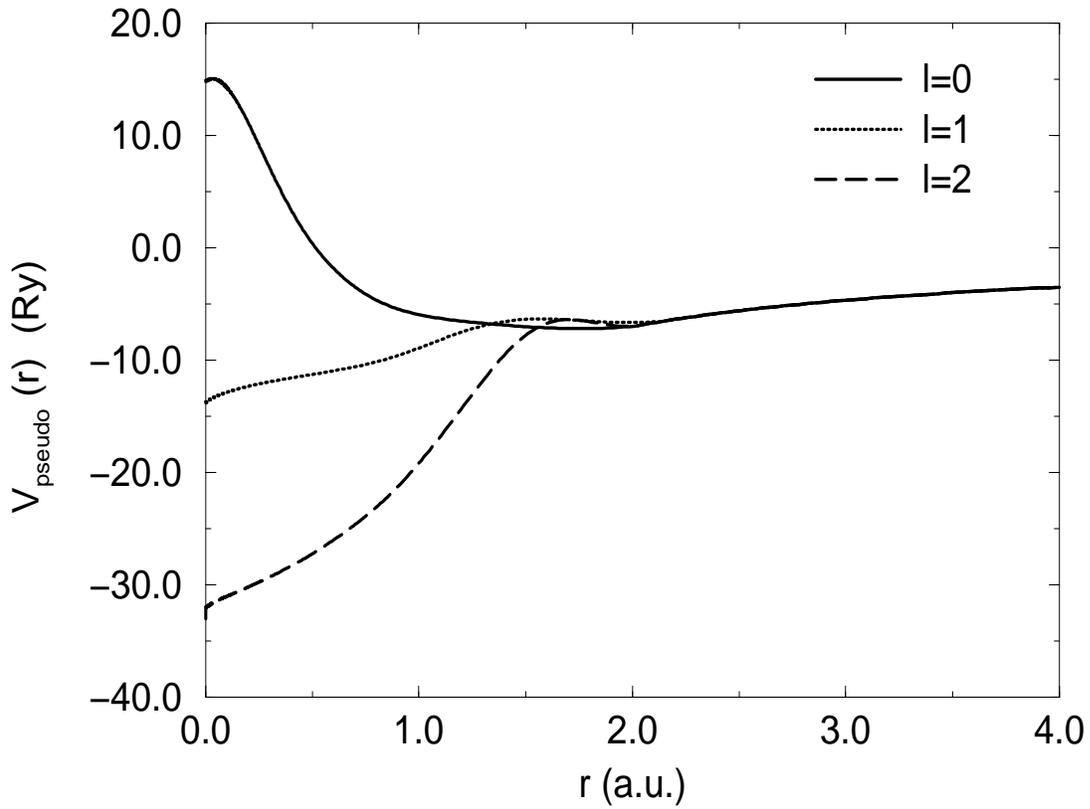,height=5.5in,width=6.0in}}
\caption{Optimized ionic pseudopotentials for Mn. The exchange-correlation 
potential contribution from the core charge has 
been subtracted, in accordance with the non-linear core
correction scheme of Louie et al. \protect \cite{NLCCs}. 
The core radii are 2.0, 2.15 and 2.0 a.u. respectively for the $s$, $p$ and $d$
pseudopotentials. Four basis functions were included in the optimization,
which used a cutoff wavevector of 7.5 a.u.}
\label{Mn_PP}
\end{figure}

\begin{figure}
\centerline{\psfig{figure=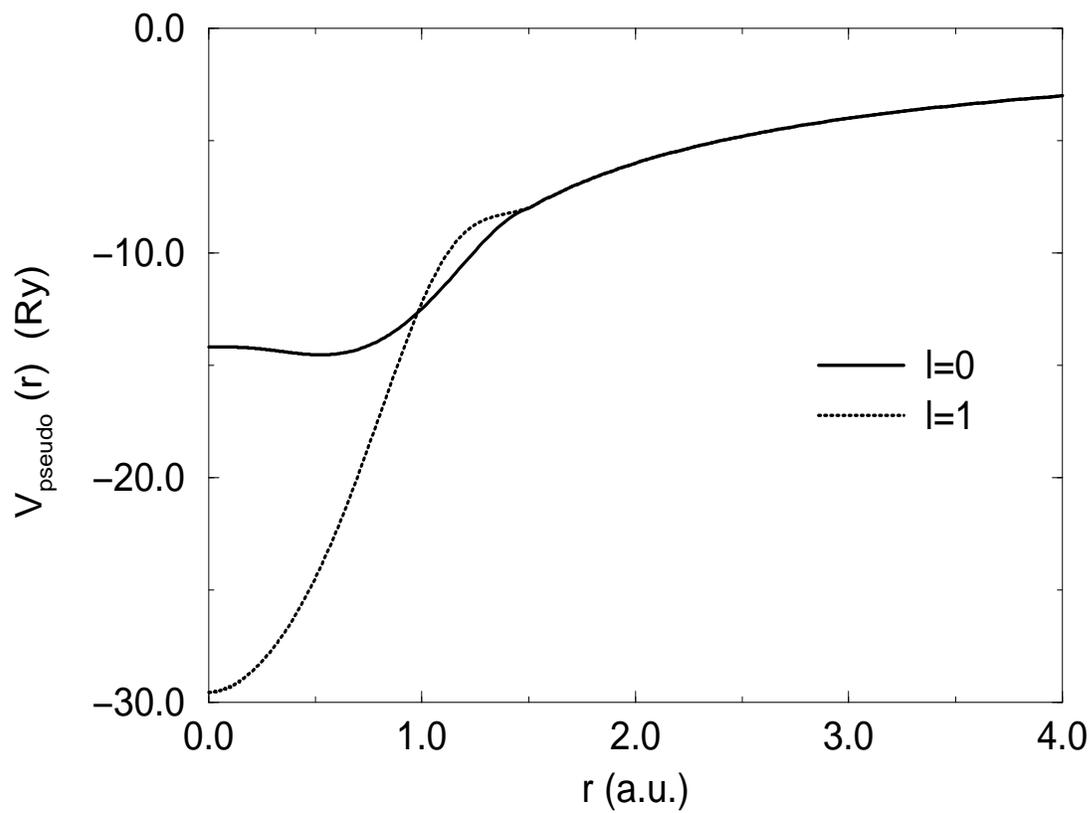,height=5.5in,width=6.0in}}
\caption{Optimized ionic pseudopotentials for oxygen.
The core radius is 1.5 a.u. The $s$ pseudopotential was optimized using 4 basis 
functions and a cutoff wave vector of 7.0 a.u., and the $p$ pseudopotential 
used 3 basis functions and a cutoff wavevector of 6.5 a.u.}
\label{Ox_PP}
\end{figure}

For Mn and O we constructed non-relativistic optimized pseudopotentials. 
The oxygen pseudopotentials were generated from a $2s^2 2p^4$ reference 
configuration with core radii of 1.5 a.u. for both $s$ and $p$ orbitals. 
They were  then optimized using 4 and 3 basis functions with 
$q_c$ of 7.0 and 6.5 a.u. for s and p orbitals respectively. 
The oxygen pseudopotentials were used in earlier calculations
for non-magnetic perovskite oxides\cite{Waghmare_Rabe} and gave 
accurate results.
For Mn, a $4s^{0.75} 4p^{0.25} 3d^{5}$ reference configuration 
was used, with  $r_c$ equal to 2.0, 2.15 and 2.0 a.u. for $s$, 
$p$ and $d$ orbitals. The optimized pseudopotentials
for Mn were constructed using 4 basis functions and $q_c = 7.5$ a.u., and
partial non-linear core corrections\cite{NLCCs} 
were included. The core charge was approximated by a zeroth order
spherical Bessel function within a radius of 0.737 a.u., with the full
core charge used outside this radius.  The resulting optimized 
pseudopotentials are shown in Figures~\ref{Mn_PP} and ~\ref{Ox_PP}.
Again the pseudopotentials are transferable to within a few meVs.

All pseudopotentials were put into separable form\cite{Kleinman_Bylander}
using two projectors for each angular momentum\cite{Blochl}.
The first projector was taken as the atomic pseudo wavefunction,
and the second projector as $r^2$ times the first projector,
suitably orthogonalized to the first\cite{Teter}.
For La, Bi and Mn, the $l=0$ component was chosen as the local potential.
For oxygen the local potential was the $l=1$ component.
The absence of ghost states was confirmed using the ghost theorem
of Gonze, K{\" a}ckell and Scheffler\cite{Gonze}.

\subsection{Technical Details}

The total energy and band structure calculations were performed
on a Silicon Graphics Power Challenge L using 
the conjugate gradient program CASTEP 2.1\cite{Payne,Castep} which we have
extended to study spin-polarized systems. We used a plane wave cut 
off of 60 Ry, which corresponds to around 3500 plane waves in a 
cubic unit cell with lattice constant of around 4 \AA\ . 
A variable Gaussian broadening between 1 eV and 0.002 eV was applied 
to the k-point sampling to speed convergence for metallic systems.
A 6x6x6 Monkhorst-Pack\cite{Monkhorst_Pack} grid was used for all
calculations. This led to 10 k-points in the irreducible Brillouin Zone
for the high symmetry cubic structures, and a correspondingly
higher number for the distorted structures with lower symmetry. 

The Perdew-Zunger parameterization\cite{Perdew_Zunger} of the 
Ceperley-Alder exchange correlation potential\cite{Ceperley_Alder}
with the von Barth-Hedin interpolation formula\cite{vonBarth_Hedin} 
was used. Previous all-electron calculations 
for perovskite manganites\cite{Pickett_Singh} found no appreciable 
difference between results obtained using the von Barth-Hedin and 
the more sophisticated Vosko-Wilk-Nusair\cite{Vosko} interpolations.

For density of states (DOS) calculations we interpolated the
calculated eigenvalues to a grid of $\approx$ 350,000 
k-points in the irreducible simple cubic Brillouin Zone using the
interpolation scheme of Monkhorst and Pack \cite{Monkhorst_Pack}. 
We then applied the 
Gilat-Raubenheimer method\cite{Gilat-Raubenheimer} to this fine 
mesh. Finally, for the band structure plots, and for use in the
tight-binding analysis, symmetry labels along high-symmetry lines 
were assigned using projection operators for the corresponding
irreducible representations.

\subsection{Comparison with LAPW calculations}

\begin{figure}
\centerline{\psfig{figure=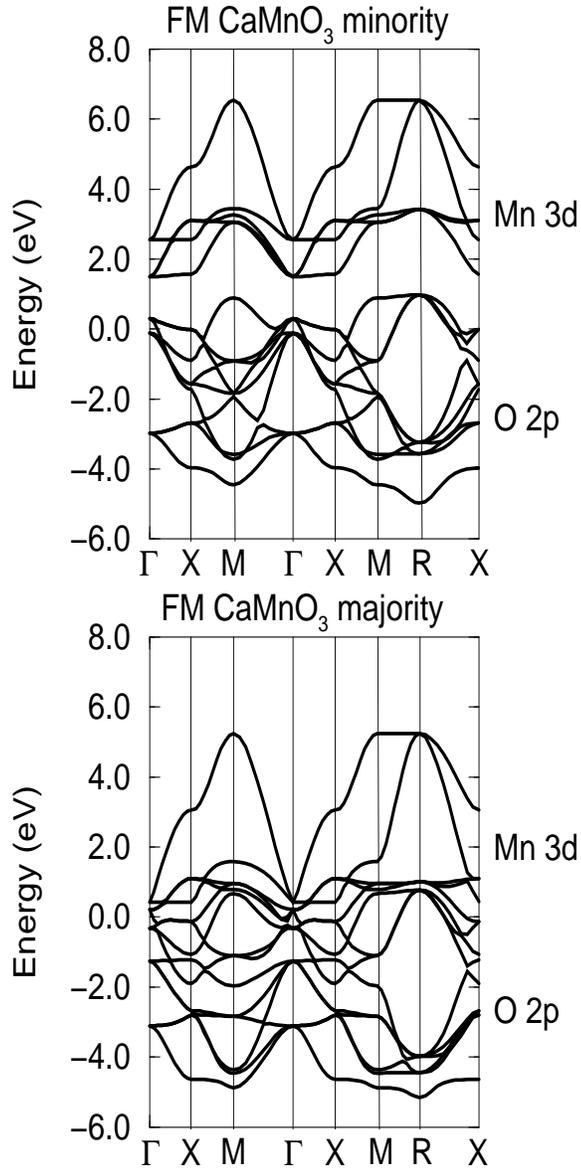,height=6.0in,width=6.0in}}
\caption{Up- and down-spin band structures for CaMnO$_3$ 
calculated using the plane wave pseudopotential method and the
partial non-linear core correction method of Louie et al.\protect\cite{NLCCs}. 
The band structures are in good
agreement with earlier LAPW calculations\protect\cite{Pickett_Singh}. }
\label{CaMnO3_NLCC}
\end{figure}

Before presenting our new results for BiMnO$_3$, we first show that 
the results obtained for perovskite manganites using this PWPP
implementation are in excellent agreement with the results of
previously published LAPW calculations. Figure~\ref{CaMnO3_NLCC}
shows the band structures for up- and down- spin electrons
in cubic ferromagnetic CaMnO$_3$, calculated in this work using
the PWPP method. 
We used a 10 electron Ca pseudopotential with a $3s^23p^53d^1$ 
Ca$^{2+}$ reference
configuration and core radii of 1.29, 1.29 and 1.45 a.u. for the $s$, $p$
and $d$ potentials. Four basis functions were included in the
optimization\cite{Rappe}, with a cutoff wavevector of 7.0 a.u.
The band structures are indistinguishable from those shown 
in Figure 7 of Ref.~\cite{Pickett_Singh}, which were calculated using 
the LAPW method. In addition, we find that our PWPP calculated energy 
difference between the ferro- and paramagnetic phases of CaMnO$_3$ is 
876 meV, within 2\% of the LAPW value of 860 meV. The inclusion of 
non-linear core corrections is essential.
The pseudopotential band structure calculated 
{\it without} non-linear core corrections is markedly different from the 
LAPW band structure. In particular the energy of the minority spin 
Mn $3d$ bands is too high, a pseudo-gap opens up in the majority spin 
O $2p$ bands, and the lower O $2p$ bands have unphysical flat regions. 
In addition the total energy difference between the ferromagnetic and
paramagnetic phases is overestimated to be 3.2 eV.

\section{Calculated Electronic Properties of BiMnO$_3$ and LaMnO$_3$}
\label{Results}

In this section we investigate the origin of the 
differences between BiMnO$_3$ and the rare earth manganites by
comparing the calculated electronic properties of BiMnO$_3$
with those of LaMnO$_3$. We begin by calculating the electronic structure
for the high symmetry cubic phases, without including magnetic effects,
then lower the magnetic symmetry to the ferromagnetic phase. Finally 
we introduce structural distortions in both paramagnetic and 
ferromagnetic calculations. 
This ability to isolate structural and magnetic
distortions is unique to computational studies,
and allows identification of the essential microscopic
interactions which cause the observed macroscopic behavior.
All calculations are carried out at the experimental lattice
constant of 3.95 \AA\ .

\subsection{Cubic Paramagnetic Structures}

First we present results for BiMnO$_3$ and LaMnO$_3$ in the
highest possible symmetry state - that is the cubic structure,
without spin polarization (we call this the paramagnetic (PM)
phase). Although this phase is experimentally inaccessible,
it provides a useful reference for understanding the 
ferromagnetic (FM) structures to be discussed later in this paper.

\begin{figure}
\centerline{\psfig{figure=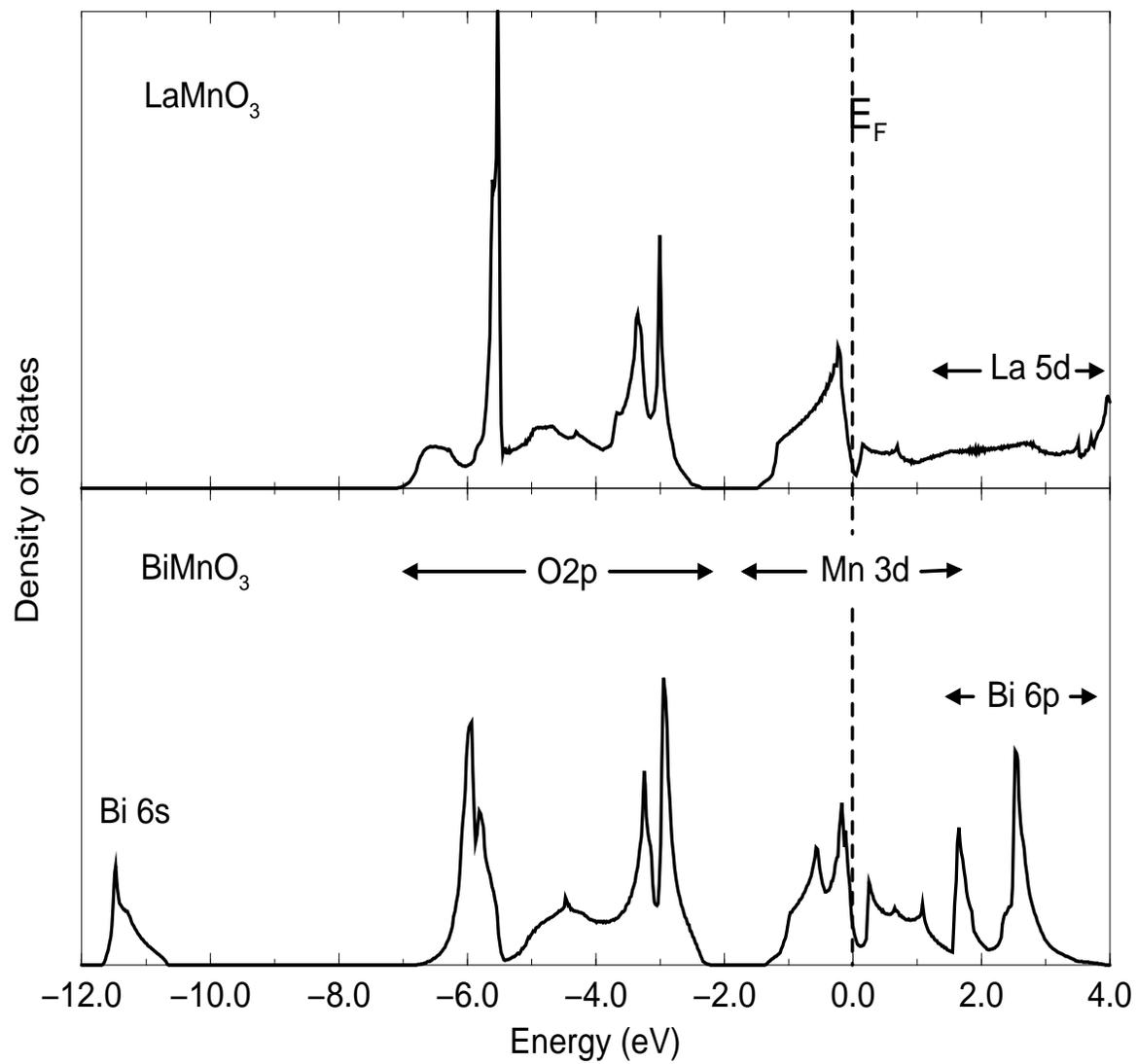,height=6.0in,width=6.5in}}
\caption{Calculated densities of states for cubic paramagnetic
LaMnO$_3$ and BiMnO$_3$.}
\label{PM_DOS}
\end{figure}

Figure~\ref{PM_DOS} shows the calculated densities of states
for cubic paramagnetic LaMnO$_3$ and BiMnO$_3$. The plotted energy
range is from -12 eV to 4 eV, and the lower lying semi-core
states have been omitted for clarity.
The Fermi level is set to zero in both cases. 
The broad series of bands between -2 and -7 eV in both materials arises
from the oxygen $2p$ orbitals. Above the oxygen $2p$ bands, and
separated from them by an energy gap, are the Mn $3d$ bands. The
Mn $3d$ bands are divided into two sub-bands - the lower energy
$t_{2g}$ bands, and the higher energy $e_g$ bands - as a result
of crystal field splitting by the octahedral oxygen anions.
One striking {\it difference} between the two DOS plots is the
presence of a band between -10 and -12 eV in the BiMnO$_3$ band
structure which does not exist in the LaMnO$_3$ case. This band
corresponds to the high lying occupied Bi $6s$ electrons. 
In addition, the high energy La $5d$ electrons have a very
different form than the Bi $6p$ electrons, which occupy a
similar energy range.

The Fermi level lies near the top of the Mn $3d$ $t_{2g}$ bands
and is in a region of high density of states.
The large DOS at the Fermi level suggests that the cubic PM structure
is unstable, and that a lower energy structure could be achieved by
allowing spin-polarization and/or structural distortion.

\begin{figure}
\centerline{\psfig{figure=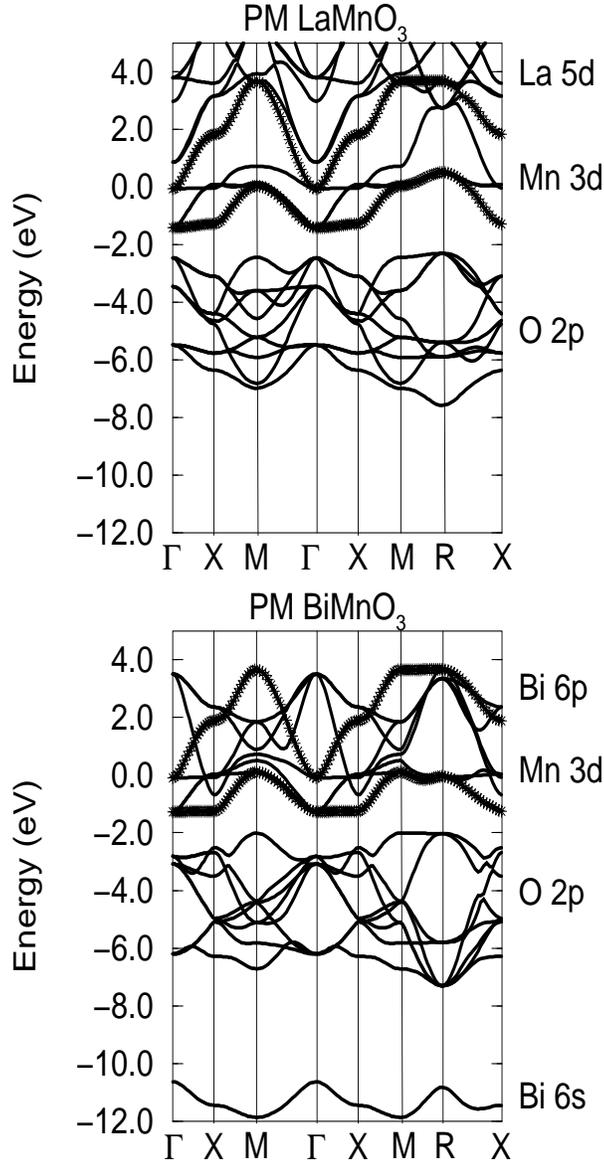,height=6.0in,width=6.0in}}
\caption{Calculated band structures for cubic paramagnetic
LaMnO$_3$ and BiMnO$_3$ along the high symmetry axes of the
Brillouin Zone.
The highlighted lines in the band
structure plots accentuate the upper and lower Mn $3d$ bands, which
have a similar form to each other and to those of CaMnO$_3$
(Figure~\ref{CaMnO3_NLCC}).}
\label{PM_BS}
\end{figure}

Figure~\ref{PM_BS} shows the corresponding band structures 
along the high symmetry axes of the simple cubic Brillouin Zone. 
Again the broad O $2p$ bands between -2 and -7 eV can be seen
clearly in both materials, with the Mn $3d$ bands above them,
separated by an energy gap. The highlighted lines in the band 
structure plots accentuate the upper and lower Mn $3d$ bands, which 
have a similar form to each other and to those of CaMnO$_3$
(Figure~\ref{CaMnO3_NLCC}). This indicates a `universality' in the 
manganite structure throughout the perovskite manganite series,
which is independent of the identity of the large cation.

Again we observe the Bi
$6s$ band, which can be seen to have considerable dispersion 
suggesting that it is involved in covalent bonding.
In addition, the Mn $3d$ bands in BiMnO$_3$ overlap with the
partially occupied Bi $6p$ orbitals, whereas in LaMnO$_3$ the next
highest bands are the unoccupied La 5$d$ bands.

\begin{figure}
\centerline{\psfig{figure=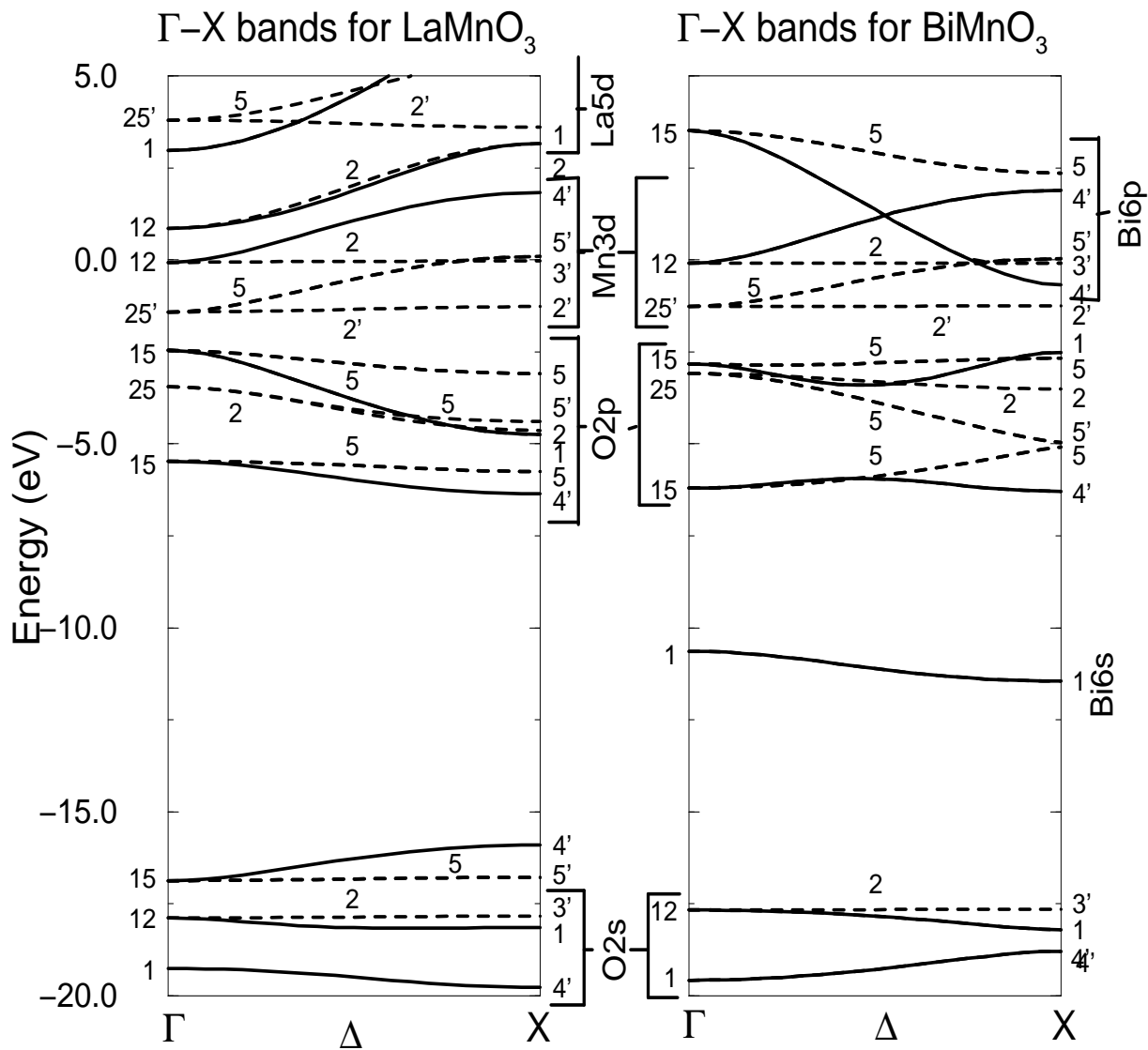,height=6.0in,width=6.5in}}
\caption{Calculated band structures for cubic paramagnetic
LaMnO$_3$ and BiMnO$_3$ along the $\Gamma \rightarrow$ X axis.
The solid lines show the bands of $\Delta_1$ symmetry.}
\label{GtoX_PM}
\end{figure}

The differences between the two band 
structures show up clearly in the bands along the $\Gamma 
\rightarrow  X$ line. This region of the band structure is shown in 
Figure~\ref{GtoX_PM} with the symmetry labels added, and the
energy scale extended to include the lower energy O $2s$ and La 5p
bands. In the following discussion we focus on the bands with
$\Delta _1$ symmetry as capturing the essential features. The
$\Delta _1$ bands 
are plotted with solid lines, and all other bands with dashed lines.

\begin{figure}
\centerline{\psfig{figure=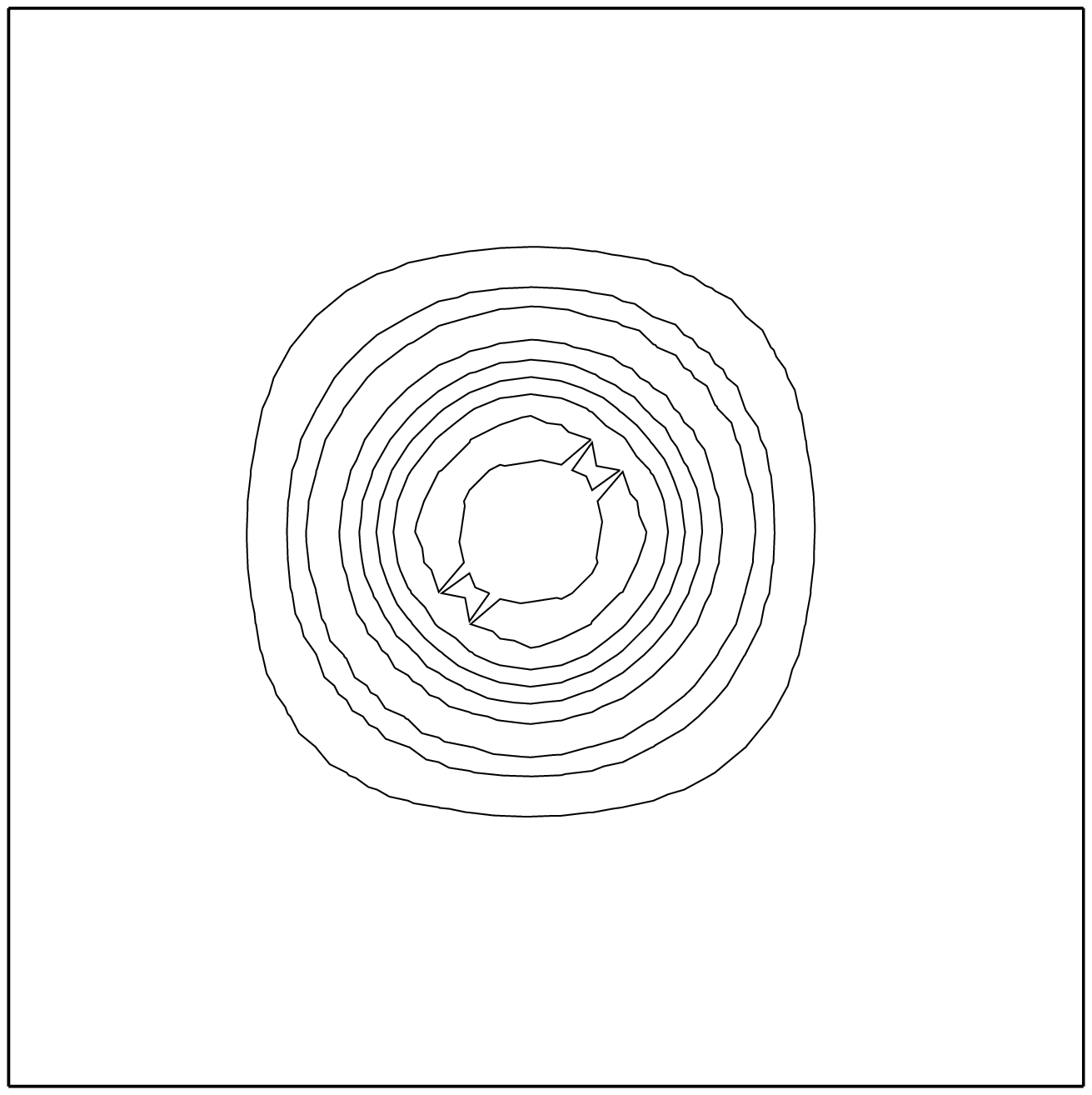,height=6in,width=6.0in}}
\caption{Contour plot of the charge density distribution in the
Mn $3d$ bands at the $\Gamma$ point of LaMnO$_3$. A (100) projection
is shown, with the Mn cation 
at the center of the square, the La ions at the corners, and
the oxygen anions at the edge mid-points and the center. The charge 
density has been integrated through the unit cell perpendicular to
the (100) plane.}
\label{LaGgnu}
\end{figure}

In LaMnO$_3$, the oxygen $2p$ $\Delta_1$ bands decrease monotonically
in energy from $\Gamma$ to X, and the Mn $3d$ $\Delta_1$ band
increases monotonically from $\Gamma$ to X. Figure~\ref{LaGgnu}
shows a contour plot of the charge density in the group of bands
above the O $2p$ manifold near the Fermi level at the $\Gamma$ point.
The spacing of the outer three lines is $\frac{1}{3}$
that of the innermost contours.
The charge density has been projected on the (100) plane by integrating
through the whole unit cell perpendicular to (100). The
Mn ion is located at the center of the plot, the La ions at the corners
and the O ions in the middle of the cell edges and at the center.
It is clear that the Fermi surface consists largely of 
Mn $3d$ electrons, with contributions from other atoms below the
resolution of the plot. The charge density at the X point is very
similar, except that there is a small oxygen contribution,
indicating that the amount of Mn $3d$ - O $2p$
hybridization increases along the $\Gamma$ to X line.

\begin{figure}
\centerline{\psfig{figure=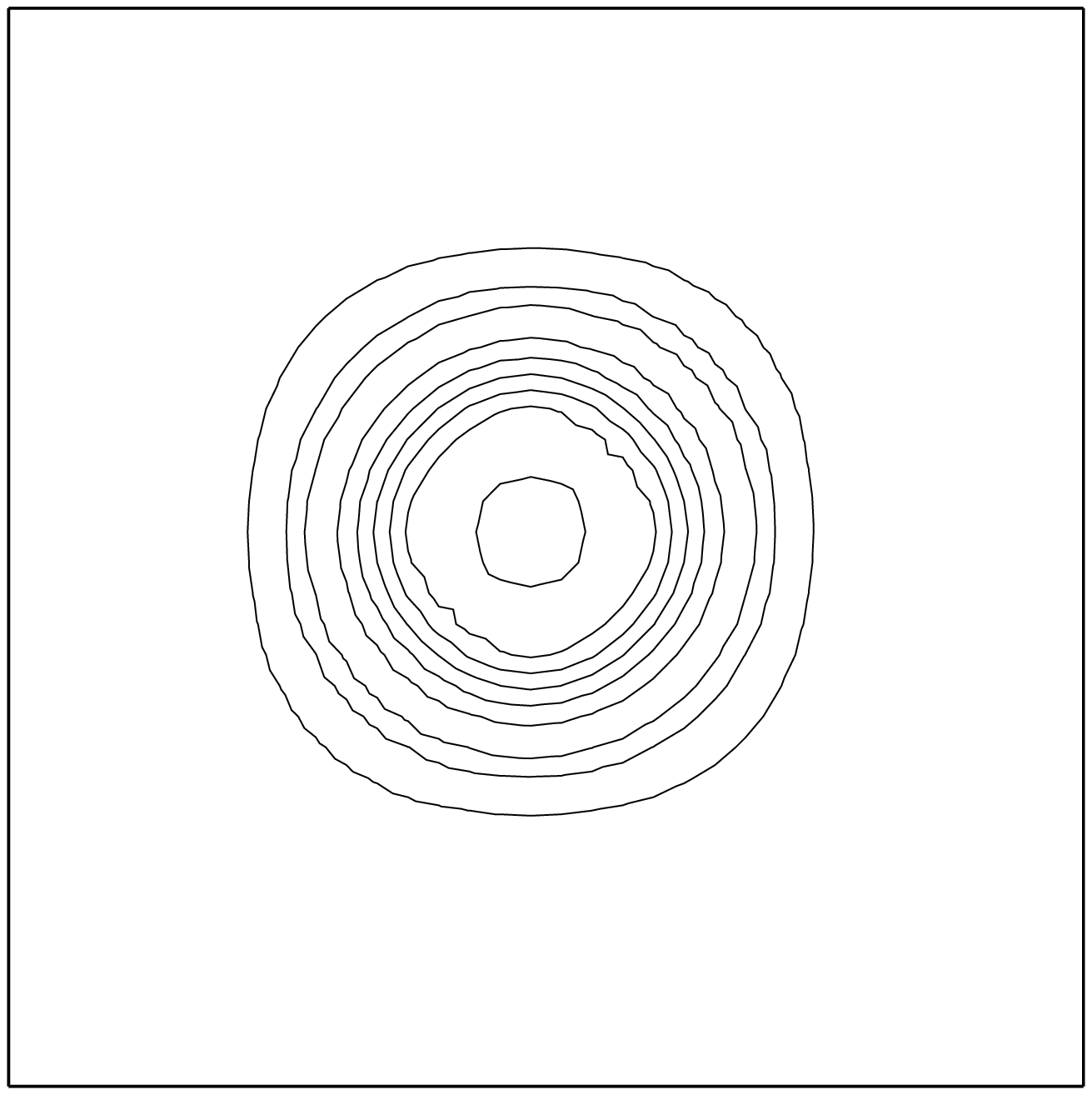,height=6in,width=6.0in}}
\caption{Contour plot of the charge density distribution in the
Mn $3d$ bands at the $\Gamma$ point of BiMnO$_3$. A (100)
projection is shown, with the Mn cation 
at the center of the square, the Bi ions at the corners, and
the oxygen anions at the edge mid-points and the center. The charge 
density has
been integrated through the unit cell perpendicular to the
(100) plane. Note the similarity to
the corresponding plot for LaMnO$_3$ (Figure~\ref{LaGgnu}).}
\label{BiGgnu}
\end{figure}

\begin{figure}
\centerline{\psfig{figure=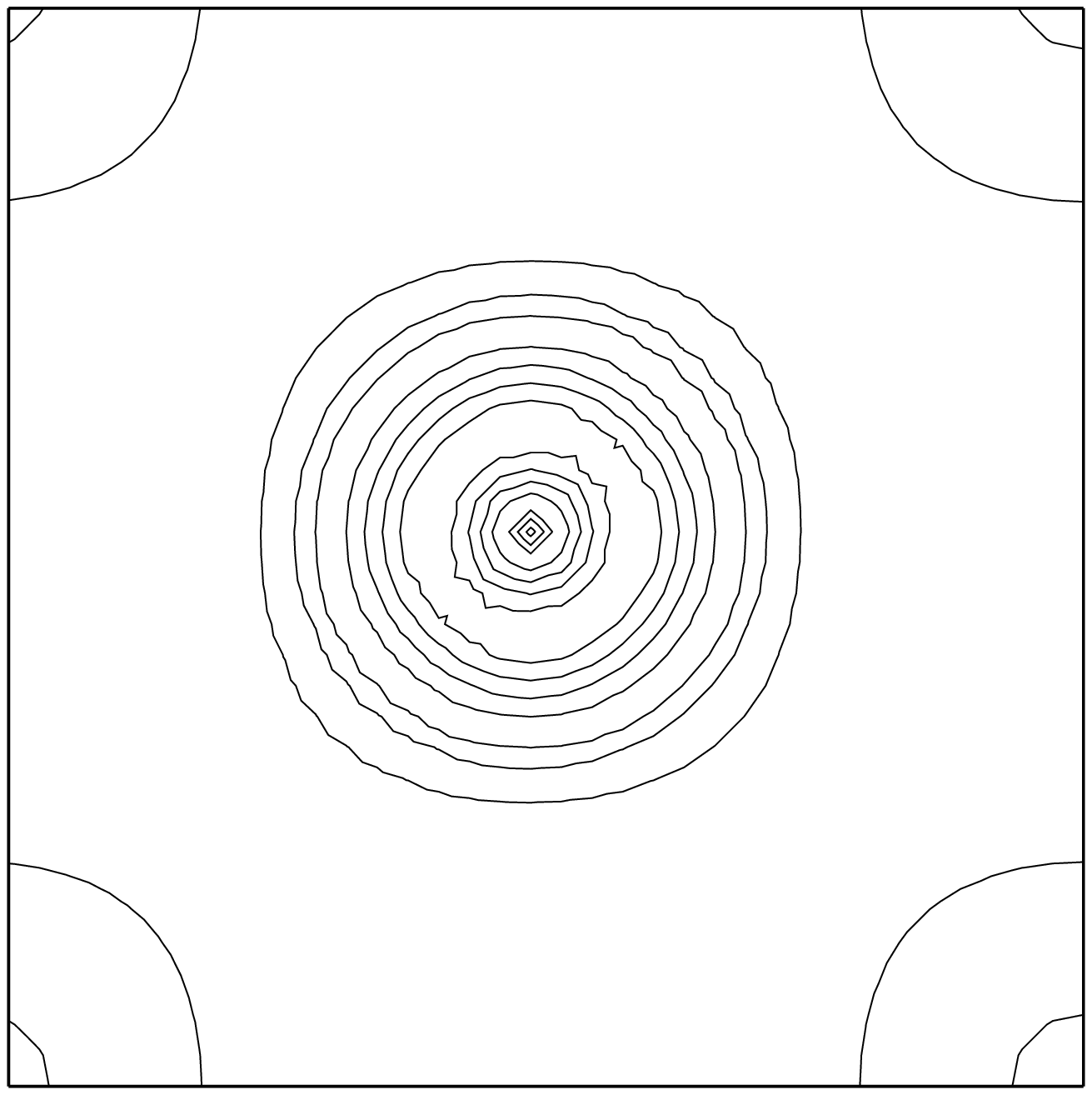,height=6in,width=6.0in}}
\caption{Contour plot of the charge density distribution in the
`Mn $3d$' bands at the X point of BiMnO$_3$. A (100)
projection is shown, with the Mn cation 
at the center of the square, the Bi ions at the corners, and
the oxygen anions at the edge mid-points and the center. The charge 
density has been integrated through the unit cell perpendicular to
the (100) plane.}
\label{BiXgnu}
\end{figure}

In BiMnO$_3$ the behavior is quite different. First, the $X_1$ 
symmetry of the Bi $6s$ band at the X-point causes the $X_1$ O 
$2p$ band to be `pushed up' in energy, resulting in a different 
ordering of the O $2p$ bands at the X point. The Mn $3d$ $\Delta_1$ 
band crosses the very dispersive Bi $6p$ $\Delta_1$ band, and the 
latter moves below the Fermi level near the X point. Figures~\ref{BiGgnu}
and~\ref{BiXgnu} show contour plots at the $\Gamma$ and X points 
respectively, of the charge density in the five bands near the Fermi level 
above the O $2p$ manifold. (These are the bands which we found to be 
entirely Mn $3d$ in the case of LaMnO$_3$.)
Again the charge density is projected onto the (100) plane,
and the same scale is used for the contours as in Fig~\ref{LaGgnu}.
At the $\Gamma$ point the charge distribution is similar to
that of LaMnO$_3$. However at X there is a significant amount of
charge density around the Bi atoms. Further analysis reveals that
the Bi component is in the $X_4'$ band, which crossed the highest
Mn $3d$ band. The lack of anti-crossing between the Bi $6p$ and
Mn $3d$ $\Delta_1$ bands is due to the zero matrix element between the 
Bi $6p$ and Mn $3d$ orbitals.

\begin{figure}
\centerline{\psfig{figure=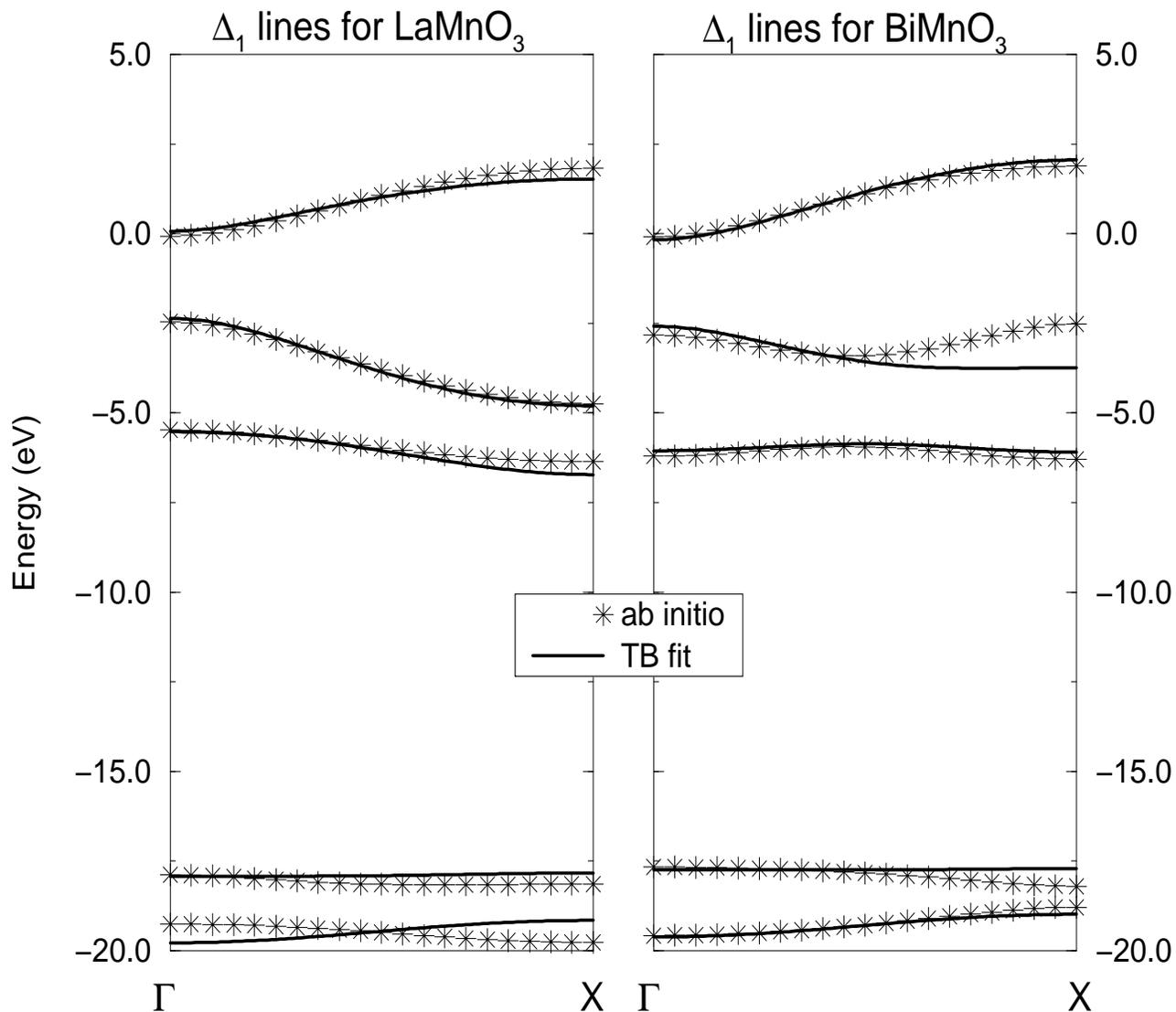,height=6in,width=6.5in}}
\caption{Comparison of {\it ab initio} $\Delta_1$ bands with those 
obtained from a tight-binding fit using only Mn and O orbitals 
in the basis. The fit for LaMnO$_3$ is good indicating that only
the Mn and O ions are significantly involved in covalent
bonding. The fit for BiMnO$_3$ has a higher RMS deviation, and
in particular misses the additional curvature around -3 eV
near the X point. This shows that additional interactions are present.}
\label{tbfit_1}
\end{figure}

To quantify the differences between cubic PM BiMnO$_3$ and LaMnO$_3$,
we performed tight-binding analyses of the $\Gamma$ to X regions of 
the respective band structures. Tight-binding parameters were obtained by 
non-linear-least-squares fitting\cite{Mattheiss} to the calculated 
{\it ab initio} energies at the high symmetry $\Gamma$ and X points, and 
at 19 points along the $\Delta$ axis.
First we performed analyses with only oxygen $2s$ and $2p$ and Mn $3d$ 
orbitals included in the basis set. The tight-binding parameters thus obtained 
are given in Table~\ref{TB_params1}. The band energies calculated
using these parameters have root mean square (RMS) deviations 
from the {\it ab initio} energies of 0.20  for LaMnO$_3$ and 0.25  for
BiMnO$_3$. The resulting band structures for the bands of $\Delta_1$
symmetry are compared with the {\it ab 
initio} values in Figure~\ref{tbfit_1}. The limited basis set
reproduces the LaMnO$_3$ bands well, consistent with an early
proposal by Goodenough\cite{Goodenough}
that the magnetic properties of LaMnO$_3$ are determined 
by the Mn $3d$ - O $2p$ hybridization only. Note that the fit to the
lower energy O $2s$ bands is the least good - these bands are
very close in energy to the La $5p$ bands which have not been included
in the fit.
The behavior of the BiMnO$_3$ $\Delta_1$ bands is less well
reproduced, confirming that additional orbital overlaps are
essential in producing the observed band structure.

\begin{figure}
\centerline{\psfig{figure=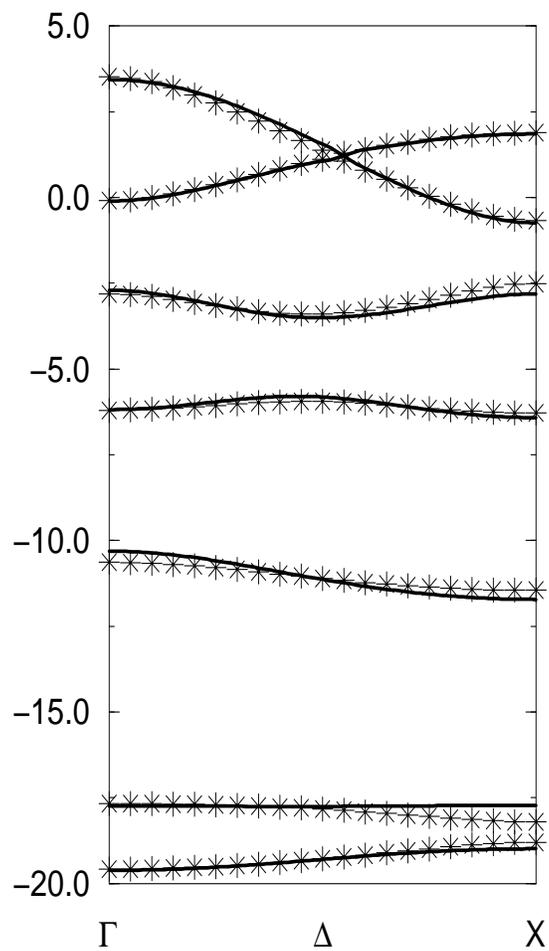,height=6.0in,width=6.0in}}
\caption{Comparison of ab initio $\Delta_1$ bands with those 
obtained from a tight-binding fit including Bi $6s$ and $6p$ orbitals
in addition to Mn $3d$ and O $2s$ and $2p$. The fit is improved over the fit
using only Mn and O orbitals.}
\label{tbfit_2}
\end{figure}

We then repeated the fitting procedure for BiMnO$_3$, with
Bi $6s$ and $6p$ orbitals added to the basis. Transfer
integrals between nearest neighbor Bi $6s$ - O $2p$, Bi $6p$ - O $2p$, 
Bi $6p$ - Mn $3d$ and Bi $6p$ - Bi $6p$ orbitals were allowed to be non-zero. 
This significantly
improved the quality of the fit to the {\it ab initio} bands (see
Figure~\ref{tbfit_2}), and reduced the root mean square deviation to 0.12.
The values of the new tight-binding parameters are given in
Table~\ref{TB_params2}. The largest transfer integrals
involving Bi are the Bi $6s$ - O $2p$ and Bi $6p$ - O $2p$ $\sigma$ 
interactions,
with the magnitude of the $\sigma$-bonded Bi $6p$ - O $2p$ interaction being 
approximately 30 \% larger than that of the Bi $6s$ - O $2p$. 
Also large are the Bi $6p$ - Bi $6p$ 
$\sigma$ interactions, which cause the Bi $6p$ $\Delta_1$ band to be
pushed down below the Fermi level.

\subsection{Comparison with Experiment}

The conclusion from our tight-binding analysis is
consistent with recent measurements showing
enhancement of charge ordering in Bi-doped CaMnO$_3$\cite{Rao}.
Rao and co-workers observed that the charge-ordered state in
Bi$_{0.3}$Ca$_{0.7}$MnO$_3$ persists to a higher temperature than
that in La$_{0.3}$Ca$_{0.7}$MnO$_3$. They explained their observations
by noting that the electronegativity of Bi enhances Bi-O hybridization
and in turn reduces the amount of Mn-O hybridization.

Additional evidence in support of this phenomenon can be extracted
from Ref. \cite{SWCheong}, in which a temperature-composition phase
diagram for Bi$_{1-x}$Ca$_x$MnO$_3$, showing the charge-ordered
transition and the N\'{e}el temperature, is plotted. Comparison
with similar data for La$_{1-x}$Ca$_x$MnO$_3$ (in for example
Ref. \cite{Schiffer}) confirms that the charge ordered
phase persists to higher temperature in Bi-doped CaMnO$_3$ than in 
La-doped CaMnO$_3$.
The ability to tune the positions of phase boundaries by
substitution of Bi for the rare earth ions might prove valuable
in optimizing materials properties for specific device applications.

\subsection{Cubic Ferromagnetic Structures}

Next, we present the results of calculations in which the high symmetry 
cubic structure is retained, but the electrons are allowed to spin polarize. 
We find that introduction of spin polarization reduces the energy of both 
BiMnO$_3$ and LaMnO$_3$ by around 1 eV per unit cell compared with the 
paramagnetic case. The most important observation of this section is that
the {\it differences} between BiMnO$_3$ and LaMnO$_3$ which we observed in 
the paramagnetic calculations persist into the ferromagnetic phase, 
the PM to FM transition introducing the same kinds of changes in both 
materials.

\begin{figure}
\centerline{\psfig{figure=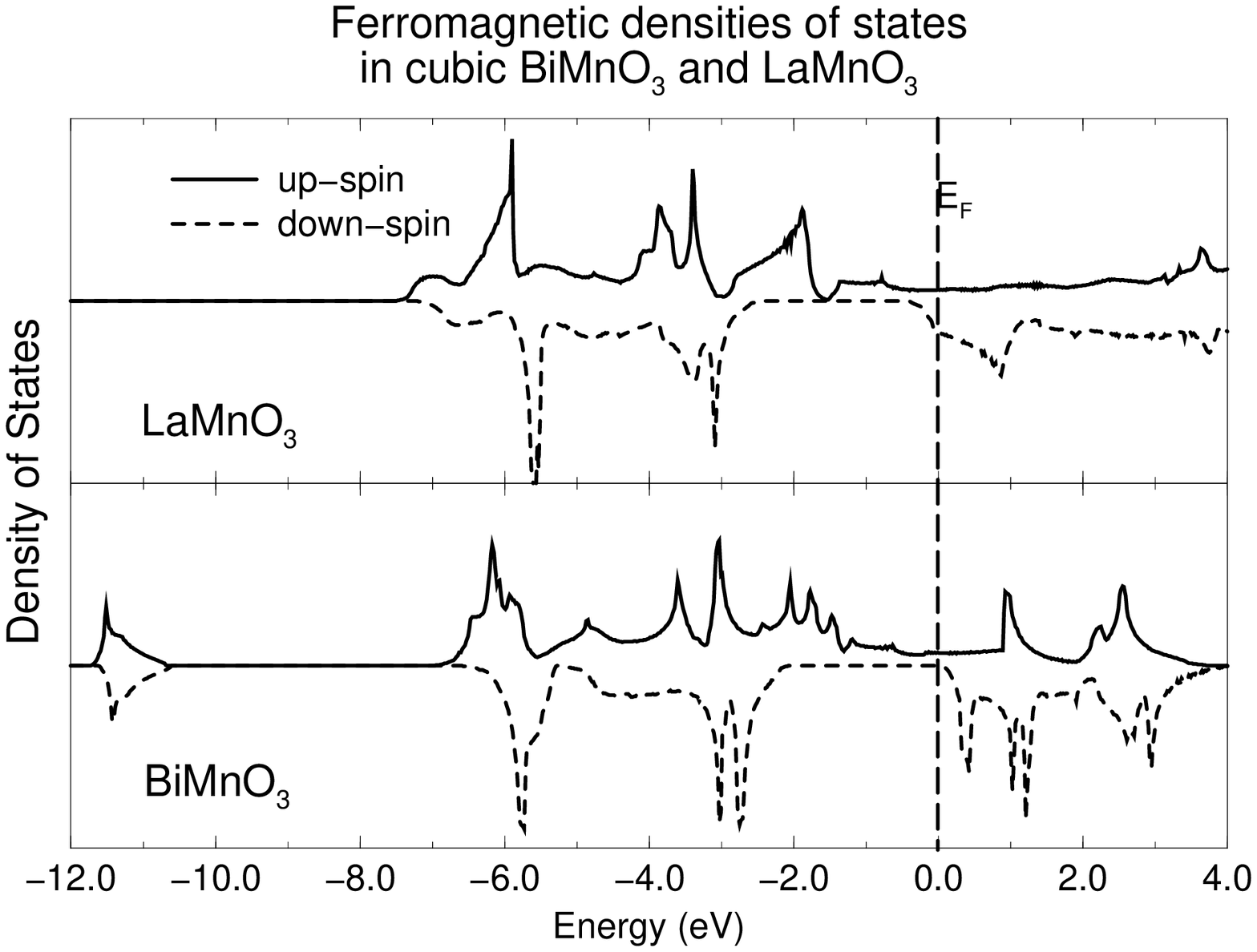,height=6.0in,width=6.0in}}
\caption{Calculated densities of states for cubic ferromagnetic LaMnO$_3$
and BiMnO$_3$.}
\label{FM_DOS}
\end{figure}

Figure~\ref{FM_DOS} shows the calculated densities of states
for cubic ferromagnetic LaMnO$_3$ and BiMnO$_3$.
The majority spins are represented by the solid
line on the positive y-axis, and the minority spins by
the dashed line on the negative y-axis. In both LaMnO$_3$ and BiMnO$_3$,
the down-spin Mn $3d$ band is split off from the O $2p$ band, and has
a similar form to the corresponding paramagnetic band. The up-spin
Mn $3d$ hybridizes strongly with the O $2p$ and there is no band
gap for the majority carriers. The up-spin DOS at the Fermi level in
LaMnO$_3$ is still high, indicating that the cubic FM state
has a high energy. This is consistent with the fact that the lowest energy 
spin polarization in {\it structurally relaxed} LaMnO$_3$ is 
antiferromagnetic\cite{Wollan_Koehler}.
The DOS at the Fermi level in BiMnO$_3$ is somewhat lower
suggesting that the FM phase should be more stable in BiMnO$_3$ than
in LaMnO$_3$.
For both compounds, the Fermi level cuts through the
very bottom of the down-spin Mn $3d$ bands, and the
conduction band is occupied almost entirely by up-spin electrons.
Materials in which the electrons at the Fermi level are 
100\% spin-polarized are
known as half-metallics, and are considered  desirable for use in
devices such as spin-transistors.

Again the most obvious differences between the two electronic
structures are the presence of the Bi 6$s$ band
between -10 and -12 eV, and the contrasting forms of the
Bi $6p$ and La $5d$ bands. 

\begin{figure}
\centerline{\psfig{figure=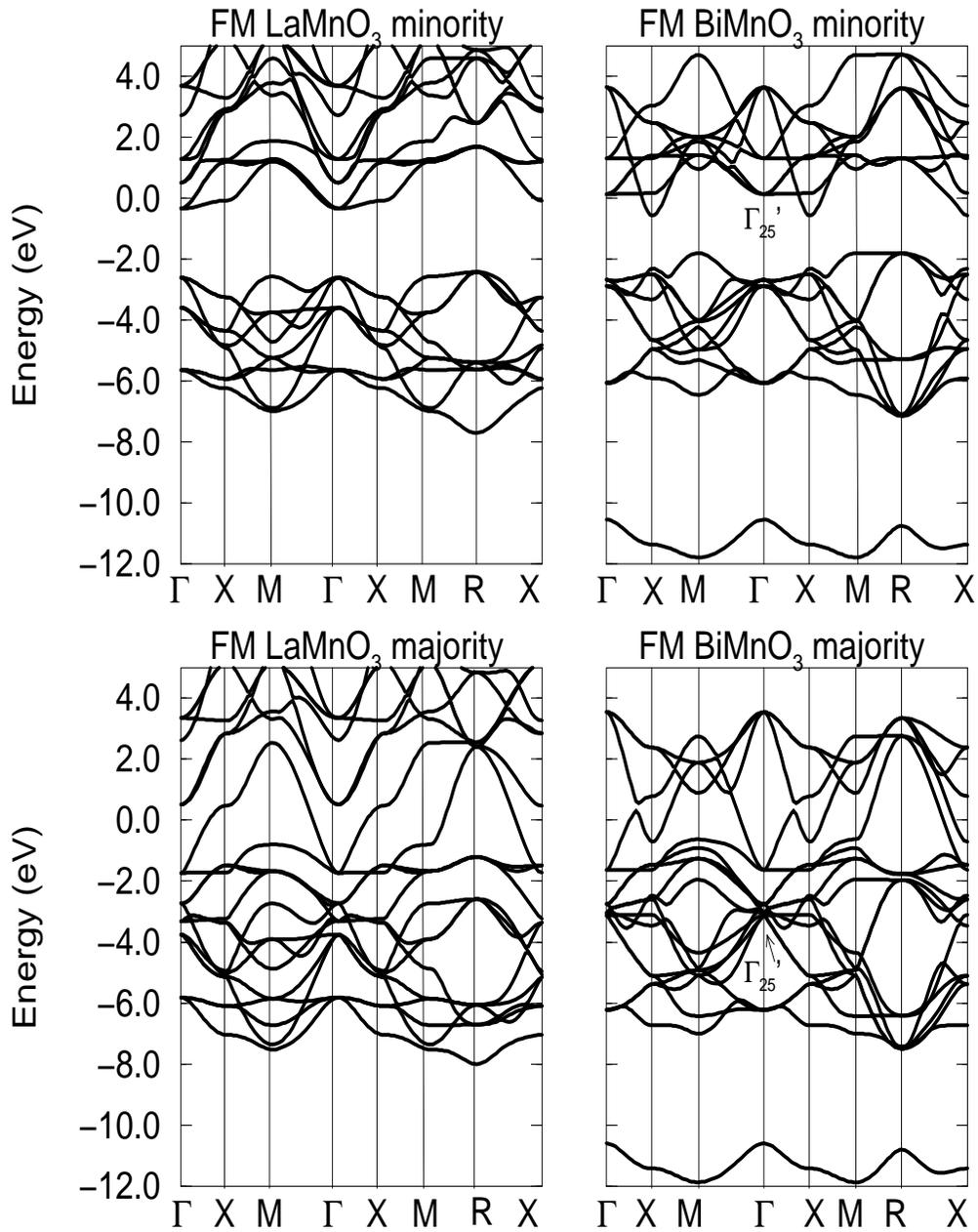,height=6.5in,width=6.5in}}
\caption{Up- and down-spin band structures for cubic LaMnO$_3$
and BiMnO$_3$ along the high symmetry axes of the 
Brillouin Zone.}
\label{FM_BS}
\end{figure}

Figure~\ref{FM_BS} shows the up and down-spin band structures
for BiMnO$_3$ and LaMnO$_3$ along the high symmetry axes of
the simple cubic Brillouin Zone. There are many similarities
between the FM and PM energy bands, and our earlier conclusions 
regarding the origins of the differences between BiMnO$_3$ and
LaMnO$_3$ continue to be valid.

First we examine the up- and down-spin LaMnO$_3$ band structures,
and compare with the PM LaMnO$_3$ band structures to determine the 
changes which spin polarization causes in a `conventional' manganite. 
The states which correspond to non-magnetic atoms are unchanged from
the paramagnetic state, and are identical for up- and down-spin
electrons. For example,
the dispersion of the lowest O $2p$ band is identical for up- and
down-spin, and for the PM phase. Also the La $5d$ bands, which
were above the Mn $3d$ bands in the PM state are unchanged in 
form and energy.
The characteristic perovskite manganite Mn $3d$ pattern
which we remarked on earlier persists in the FM phase, appearing
around 2 eV higher for the down-spin electrons than for the 
up-spin because of the exchange splitting. The up-spin Mn $3d$ and
O $2p$ bands are strongly hybridized and there is no gap between them.
However, the down-spin Mn $3d$ are split off from the O $2p$ bands
by a larger gap than in the PM case. As a result, the Mn $3d$ bands 
occupy the same energy region as the unoccupied La $5d$ bands. 

Next we compare the up- and down-spin BiMnO$_3$ band structures.
As expected, the non-magnetic Bi $6s$ band and the lower O $2p$ bands 
are identical for up- and down- spin. In fact the overall form of 
the {\it down-spin} O $2p$ bands is almost identical to that of the 
corresponding PM BiMnO$_3$ bands. However it is different from that
of the down spin LaMnO$_3$ O $2p$ bands - in particular the structure 
at the $\Gamma$ point is quite different, a consequence of
the unusual behavior of the $\Delta_1$ band which we 
observed earlier in the PM bands. Both up- and down-spin
Mn $3d$ bands show the characteristic perovskite manganite
features (most noticeable in this case along the M-R-X directions) which we 
noted earlier in our discussions of the PM band structures, and of CaMnO$_3$.
The form of the down-spin Mn $3d$/Bi $6p$ bands is 
similar to that of the
PM phase, although the energy separation from the O $2p$ bands is
larger.
In contrast, the majority bands show strong hybridization
between the Mn $3d$ and O $2p$ electrons.
The lowest Mn $3d$ bands
at the $\Gamma$ point ($\Gamma_{25}^{'}$) move below the upper
O $2p$ bands. At the R point,
the Bi $6p$ bands are now above the Mn $3d$ bands, whereas in the
down-spin and paramagnetic band structures they were below.

\begin{figure}
\centerline{\psfig{figure=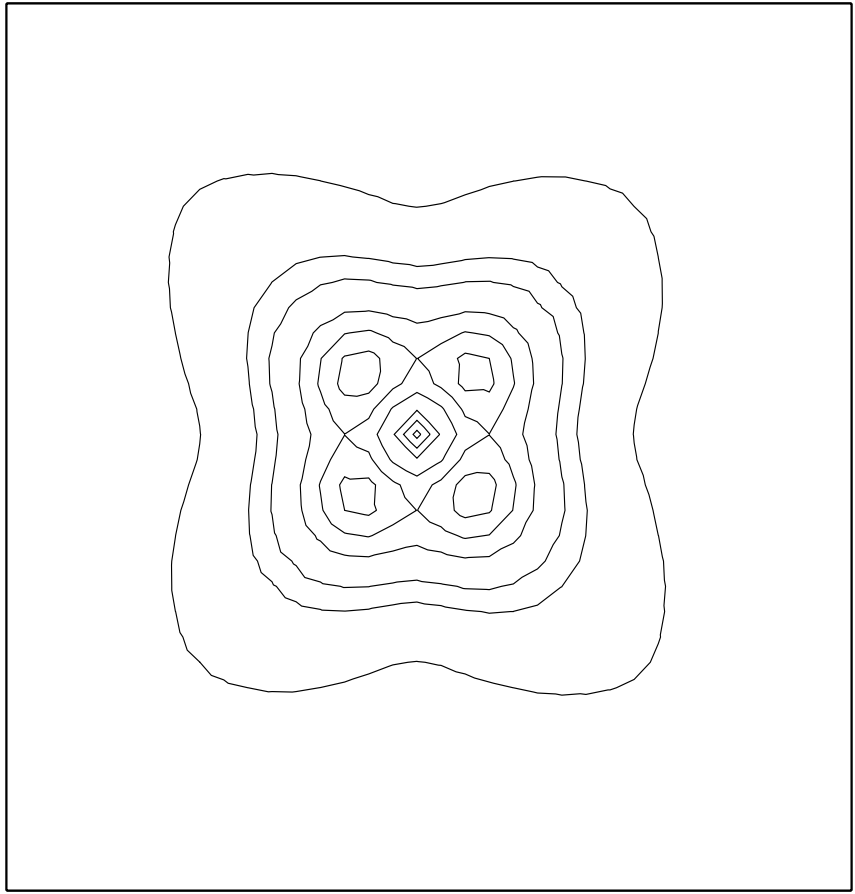,height=6in,width=6.0in}}
\caption{Minority conduction band charge density in cubic FM
LaMnO$_3$. The charge density is projected on the (001) plane
of the cubic unit cell.}
\label{FMLagnu}
\end{figure}

\begin{figure}
\centerline{\psfig{figure=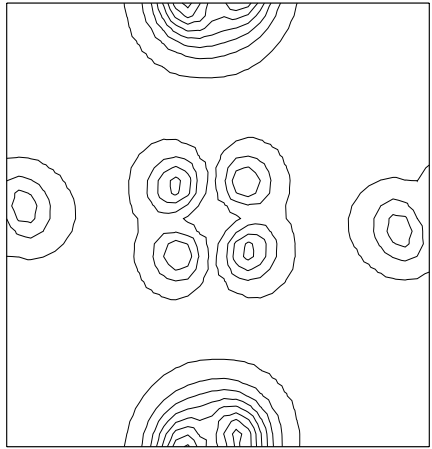,height=3in,width=3.0in}}
\centerline{\psfig{figure=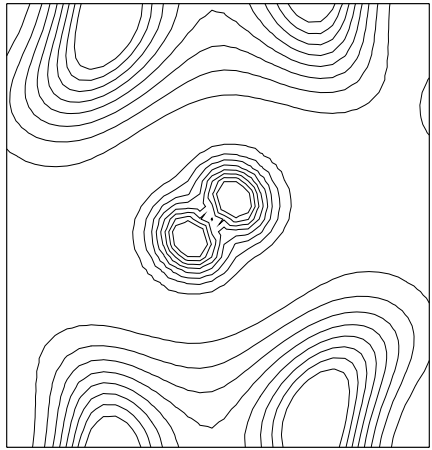,height=3in,width=3.0in}}
\caption{Minority conduction band charge density in cubic FM
BiMnO$_3$. The upper plot is a (100) slice through the Mn-O plane,
and the lower plot is a (100) slice through the Bi-O plane.}
\label{FMBignu}
\end{figure}

The contrast between the BiMnO$_3$ and LaMnO$_3$ minority 
conduction electrons is striking. Figure~\ref{FMLagnu} is a
contour plot of the small amount of down spin charge density
in the region above, and separated by a gap from, the O $2p$ bands. The charge
density through the entire unit cell has been projected onto the
(100) plane. The electrons
are localized entirely in the Mn $3d$ orbitals with no 
hybridization with the O atoms. This is a result of the
large energy separation between the Mn and O bands, which in
turn is a consequence of the minority Mn $3d$ bands being pushed to
higher energy by 
exchange forces. There is no contribution from the
La atoms, because the La $5d$ electrons never intersect the
Fermi level. Figure~\ref{FMBignu} shows the charge density of the
down-spin conduction band electrons in BiMnO$_3$. Rather than
summing over the unit cell, two different slices
through the unit cell are shown. The upper plot is in the Mn-O (100) plane,
with the Mn ion at the center and O ions at the edge mid-points.
There is minority spin conduction band charge density localized on
both the Mn and O ions.
The lower plot is in the Bi-O (100) plane, with Bi ions
at the corners and an O ion at the center.
We see that the Bi contribution to the charge density is
both large and highly directional.
The fact that the conduction electrons partly occupy $p$-type
atomic orbitals should produce quite different transport characteristics
than those observed in conventional rare earth manganites, where
the conduction bands are entirely Mn $3d$-type.

\subsection{Ferroelectric distortions}

Finally we investigate the origin of the proposed ferroelectricity
in BiMnO$_3$. In keeping with the philosophy of this paper, we
study the lattice distortions of the high symmetry cubic phases, 
and compare our results for BiMnO$_3$ with those for
LaMnO$_3$. We restrict our discussion to zone center phonons.

The perovskite manganites have five atoms per unit cell, which 
results in 15 phonon branches, 3 acoustical and 12 optical. At 
the $\Gamma$ point all phonons are three-fold degenerate, so there
is one acoustical phonon frequency (which is zero), and four
optical frequencies. We are interested in the optical phonons 
which have negative eigenvalues, indicating lattice instabilities.

The force constant matrices for LaMnO$_3$ and BiMnO$_3$ were determined
by calculating the Hellmann-Feynman forces resulting from the 
displacement of each atom in turn 0.1 \AA\ along the z direction
of the unit cell.
The forces exerted on the Mn ions by the other ions were determined using 
the acoustic sum rule. We calculated the Mn-Mn 
force by applying the acoustic sum rule to both the columns
and the rows of the resulting matrix. The two values
differed by less than around 0.001 $\frac{eV}{\mbox{\AA}}$. 
Our calculated force constant matrices for paramagnetic 
LaMnO$_3$ and BiMnO$_3$ are tabulated in Table~\ref{fcmatrices}.

\begin{figure}
\centerline{\psfig{figure=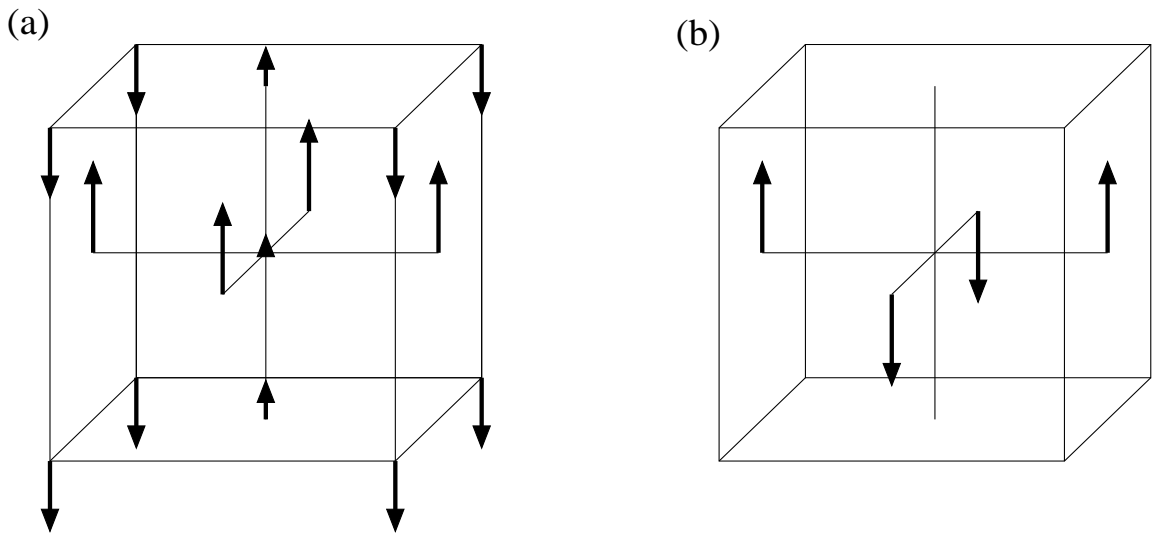,height=2.8in,width=6.0in}}
\caption{Eigenvectors of the two unstable $\Gamma$ point phonon modes in
BiMnO$_3$. The Mn ion is at the center of the unit cell surrounded by
an octahedra of oxygens, with the Bi ions at the unit cell corners. Each
mode is three-fold degenerate. (a) is the mode which leads to a 
ferroelectric distortion; (b) is a non-ferroelectric mode.}
\label{phonon}
\end{figure}

The paramagnetic cubic phases of both LaMnO$_3$ and BiMnO$_3$
have two unstable phonon modes. The frequencies and
eigenvectors of the unstable modes are given in Table~\ref{phonons}.
The mode involving motion of the equatorial oxygens, shown in
Figure~\ref{phonon} (b), does not correspond
to a ferroelectric distortion and will not be discussed further. The 
displacement pattern for the
second unstable mode is shown in Figure~\ref{phonon} (a). Here 
the large (Bi or La) cations
are moving in opposition to the oxygen cage, resulting in a ferroelectric
displacement. The imaginary frequency of this mode is twice as large
in BiMnO$_3$ than in LaMnO$_3$, indicating a stronger instability in
the Bi compound. 
It is interesting to note that the Mn is moving in the same
direction as the oxygen ions. This is opposite to the behavior
of the Ti ion in BaTiO$_3$ and PbTiO$_3$, but similar to the behavior
of Zr in PbZrO$_3$\cite{ghosez}.

The imaginary frequency phonons in the ferromagnetic phases are less 
unstable than the corresponding imaginary frequency paramagnetic phonons. 
In fact the ferroelectric mode in LaMnO$_3$ is now only slightly unstable, 
at $21.1i$  $cm^{-1}$, and the other formerly
unstable mode now has a positive frequency. 
However the frequency of the ferroelectric mode in
ferromagnetic BiMnO$_3$ remains strongly unstable, at $82.30i$ $cm^{-1}$.

It is clear from our analysis, that the presence of ferroelectricity in 
BiMnO$_3$, and absence of ferroelectricity in LaMnO$_3$, can be explained
by the different zone center lattice instabilities. Although a 
definitive prediction requires calculation of the phonon dispersion
throughout the entire Brillouin Zone, it is likely that the weakly
unstable zone center phonon in LaMnO$_3$ will be overshadowed by
a stronger instability elsewhere in the Brillouin zone, reproducing
theoretically 
the experimentally observed Jahn-Teller distortion. Similarly,
the very unstable ferroelectric mode at the zone center in BiMnO$_3$
is likely to dominate over possible unstable modes at other
frequencies, confirming theoretically the suggested existence of 
ferroelectricity in BiMnO$_3$.
We are in the process of developing spin-polarized
density functional perturbation theory code\cite{GonzePT} which will allow 
us to investigate efficiently the nature of the phonons throughout the
entire Brillouin zone. Our results will be the subject of a future
publication.

An alternative suggestion for the origin of the ferroelectricity in BiMnO$_3$ 
is that it originates from an electronic phase transition, in contrast to
the conventional displacive ferroelectricity caused by a
lattice distortion\cite{LuPC}. Such {\it electronic ferroelectricity} has been
demonstrated theoretically\cite{Portengen} in the insulating
phase of the Falicov-Kimball model, and there is some experimental data
which is consistent with the occurrence of electronic 
ferroelectricity\cite{Portengen}. 
The correlations associated with the electronic instability are included via a  BCS-like wavefunction and thus the resulting energy is not expected to be well  described within the local density approximation. 
In similar situations, however, LDA results have been successfully used to  
compute values of model parameters for input to many-body 
investigations\cite{Hybertsen}.
This is clearly a subject which 
merits further experimental and theoretical study.

\section{Summary}
\label{Summary}

Our results indicate that covalent bonding between the bismuth
cations and oxygen anions in BiMnO$_3$ introduces additional orbital 
interactions compared with the rare earth manganites, in which the rare earth -
oxygen interaction is essentially purely ionic. These additional orbital
interactions in turn stabilize different magnetic and structural phases. 
Our results are consistent with the limited experimental data 
on doped BiMnO$_3$-based compounds, and suggest that a modern 
experimental study of pure BiMnO$_3$ might reveal novel and 
potentially useful phenomena.

\section{Acknowledgments}

The authors acknowledge many useful discussions with 
Umesh Waghmare, Eric Cockayne, Serdar \"{O}\u{g}\"{u}t and Steven Lewis.
We thank the authors of Castep for providing us with their
software package, and Umesh Waghmare for implementing the
extension to metallic systems.
Funding for this work was provided by Hewlett Packard Laboratories
and by the Office of Naval Research, contract number N00014-91-J-1247.
We thank Profs. C.N.R. Rao and Tony Cheetham for sharing their
experimental results prior to publication, and Prof. Lu Sham and Dr.
Morrel Cohen for
bringing electronic ferroelectricity to our attention.
Many of the ideas presented in this manuscript were formulated 
during a workshop at the Aspen Center for Physics.

\clearpage

%
%

\begin{table}
\begin{center}
\begin{tabular}{|| l | r | r ||} \hline
                           &  BiMnO$_3$ &  LaMnO$_3$  \\ \hline
$E_{O2s}$                  & -17.680595  & -17.836376  \\
$E_{O2p}$                  &  -3.444399  &  -4.514022  \\
$E_{Mn3d}$                 &  -1.202430  &  -1.191285  \\
$V_{O2s-O2s}$              &  -0.252108  &  -0.243683  \\
$V_{(O2p-O2p)\sigma}$      &   0.735211  &   0.620568  \\
$V_{(O2p-O2p)\pi}$         &  -0.132201  &  -0.063466  \\
$[V_{(O2p-O2p)\sigma}]_2$  &  -0.211935  &   0.182635  \\
$[V_{(O2p-O2p)\pi}]_2$     &  -0.044754  &   0.082951  \\
$V_{O2s-Mn3d}$             &  -1.720688  &  -1.735814  \\
$V_{(O2p-Mn3d)\sigma}$     &  -1.964201  &  -1.838490  \\
$V_{(O2p-Mn3d)\pi}$        &   1.036316  &   0.878961  \\
$V_{(Mn3d-Mn3d)\delta}$    &  -0.003391  &   0.066346  \\ \hline
\end{tabular}
\end{center}
\caption{Tight-binding parameters for BiMnO$_3$ and LaMnO$_3$
obtained by non-linear-least-squares fitting to the {\it ab initio}
eigenvalues along $\Gamma$ to X. E indicates an orbital energy,
and V an inter-atomic transfer integral. All transfer integrals
are between nearest neighbors, except those with the subscript `2'
which are between next nearest neighbors. Only the parameters listed 
in the table were allowed to be non-zero in the fitting procedure.}
\label{TB_params1}
\end{table}

\begin{table}[p]
\begin{center}
\begin{tabular}{|| l | r ||} \hline
                      &   BiMnO$_3$   \\ \hline
$E_{Bi6s}$            &  -10.310688  \\
$E_{Bi6p}$            &    0.202695  \\
$E_{O2s}$             &  -17.717958  \\
$E_{O2p}$             &   -3.725773  \\
$E_{Mn3d}$            &   -1.167871  \\
$V_{Bi6p-Bi6p\sigma}$ &    0.848443  \\
$V_{Bi6p-Bi6p\pi}$    &    0.166061  \\
$V_{Bi6s-O2p}$        &   -0.812502  \\
$V_{(Bi6p-O2p)\sigma}$  &   -1.061660  \\
$V_{(Bi6p-O2p)\pi}$     &   -0.145201  \\
$V_{(Bi6p-Mn3d)\sigma}$ &    0.170850  \\
$V_{(Bi6p-Mn3d)\pi}$    &    0.078144  \\
$V_{O2s-O2s}$           &   -0.237506  \\
$V_{(O2p-O2p)\sigma}$   &    0.652059  \\
$V_{(O2p-O2p)\pi}$      &   -0.125450  \\
$[V_{(O2p-O2p)\sigma}]_2$  &   -0.010196\\
$[V_{(O2p-O2p)\pi}]_2   $  &   -0.002797\\
$V_{O2s-Mn3d}$          &   -1.645508  \\
$V_{(O2p-Mn3d)\sigma}$  &   -1.926533  \\
$V_{(O2p-Mn3d)\pi}$     &    0.959768  \\
$V_{(Mn3d-Mn3d)\delta}$ &    0.029347  \\ \hline
\end{tabular}
\end{center}
\caption{Tight-binding parameters for BiMnO$_3$
obtained by non-linear-least-squares fitting to the {\it ab initio}
eigenvalues along $\Gamma$ to X. An expanded tight-binding basis
including the Bi $6s$ and $6p$ orbitals was used. The labeling scheme 
for the parameters is
the same as in Table~\ref{TB_params1}. Only the parameters shown were
allowed to be non-zero.}
\label{TB_params2}
\end{table}

\begin{table}[p]
\begin{center}
\begin{tabular}{|c|r|r|r|r|r||c|r|r|r|r|r|}
  \multicolumn{6}{c}{BiMnO$_3$} & \multicolumn{6}{c}{LaMnO$_3$}     \\ \hline
   &  Bi   &  Mn   &   Oz  &   Ox  &  Oy   &
   &  La   &  Mn   &   Oz  &   Ox  &  Oy   \\ \hline
Bi & -5.47 & -2.37 & -0.73 &  4.28 &  4.28 &
La & -0.05 & -2.64 & -0.71 &  1.70 &  1.70 \\
Mn & -2.46 & 11.12 & -8.66 &  0.00 &  0.00 &
Mn & -3.07 & 15.10 &-11.57 & -0.21 & -0.21 \\
Oz & -0.64 & -8.57 & 13.03 & -1.91 & -1.91 &
Oz & -0.72 &-11.53 & 17.04 & -2.40 & -2.40 \\ 
Ox &  4.28 & -0.09 & -1.82 & -2.19 & -0.18 &
Ox &  1.92 & -0.46 & -2.38 & -0.01 &  0.91 \\ 
Oy &  4.28 & -0.09 & -1.82 & -0.18 & -2.19 &
Oy &  1.92 & -0.46 & -2.38 &  0.91 & -0.01 \\ \hline
\end{tabular}
\end{center}
\caption{Force constant matrices 
in cubic paramagnetic BiMnO$_3$ and LaMnO$_3$.}
\label{fcmatrices}
\end{table}

\begin{table}[p]
\begin{center}
\begin{tabular}{||c|c|c|c|c|c||c|c|c|c|c|c||} 
  \multicolumn{6}{c}{BiMnO$_3$}     &   \multicolumn{6}{c}{LaMnO$_3$} \\ \hline
$\nu$ (cm$^{-1})$ &  Bi  &  Mn  &  Oz  &   Ox  &  Oy  &   
$\nu$ (cm$^{-1})$ &  La  &  Mn  &  Oz  &   Ox  &  Oy   \\ \hline
$72.39i$ & 0.0 & 0.0 & 0.0 & $-\frac{1}{\sqrt{2}}$ &  $ \frac{1}{\sqrt{2}}$ & 
$49.04i$ & 0.0 & 0.0 & 0.0 & $-\frac{1}{\sqrt{2}}$ &  $ \frac{1}{\sqrt{2}}$  \\
$98.20i$ & -0.43 & 0.09 & 0.16 & 0.62 & 0.62 & 
$44.69i$ & -0.59 & 0.22 & 0.21 & 0.53 & 0.53  \\ \hline
\end{tabular}
\end{center}
\caption{Eigenvectors and eigenvalues of the 
dynamical matrix  which correspond to the unstable phonon modes
in cubic paramagnetic BiMnO$_3$ and LaMnO$_3$.}
\label{phonons}
\end{table}

\clearpage
%
%
                 
\clearpage

\end{document}